\documentclass[11pt]{article}

\usepackage[final]{acl}

\usepackage{times}
\usepackage{latexsym}

\usepackage[T1]{fontenc}

\usepackage[utf8]{inputenc}

\usepackage{microtype}

\usepackage{inconsolata}

\usepackage{graphicx}
\usepackage{xspace}
\usepackage[dvipsnames]{xcolor}
\usepackage{booktabs}
\usepackage{multirow} 

\usepackage[most]{tcolorbox}
\usepackage{listings}
\tcbuselibrary{listings}
\usepackage{float}
\usepackage{cleveref}

\definecolor{jsonstr}{HTML}{0B7C5A}
\definecolor{jsonkw}{HTML}{B5267A}

\usepackage{subcaption}

\newcommand{\marquis}{\textsc{MARQUIS}\xspace}
\newcommand{\cag}{\textsc{CAG}\xspace}
\newcommand{\ginger}{\textsc{GINGER}\xspace}
\newcommand{\rlm}{\textsc{RLM}\xspace}
\newcommand{\rootlm}{Root LM\xspace}
\newcommand{\ssqabase}{\textsc{SS-QA-Base}\xspace}
\newcommand{\ssqaginger}{\textsc{SS-QA-\ginger}\xspace}
\newcommand{\iterqabase}{\textsc{Iter-QA-Base}\xspace}
\newcommand{\iterqaginger}{\textsc{Iter-QA-\ginger}\xspace}
\newcommand{\bulletmethod}{\textsc{Bullet}\xspace}
\newcommand{\rankvideo}{\textsc{RankVideo}\xspace}

\title{\marquis: A Three-Stage Pipeline for Video Retrieval-Augmented Generation}

\author{
\quad Debashish Chakraborty\Thanks{Equal Contribution}\textsuperscript{2}
\quad Dengjia Zhang\footnotemark[1]\textsuperscript{1}
\quad Jialiang Jin\footnotemark[1]\textsuperscript{1} \\
\quad \textbf{Hanting Liu\footnotemark[1]\textsuperscript{1}}
\quad \textbf{Katherine Guerrerio\footnotemark[1]\textsuperscript{1}}
\quad \textbf{Hanxiang Qin\textsuperscript{1}}
\quad \textbf{Tyler Skow\textsuperscript{1}}  \\
\quad \textbf{Alexander Martin\textsuperscript{1}}
\quad \textbf{Reno Kriz\textsuperscript{1,2}}
\quad \textbf{Benjamin Van Durme\textsuperscript{1,2}}
\\
  \textsuperscript{1}Johns Hopkins University\quad \\\textsuperscript{2}Human Language Technology Center of Excellence
  \\
  \texttt{\small{\{dchakra6, amart233\}@jhu.edu}}
}

\begin{document}
\maketitle

\begin{abstract}

Retrieval-augmented generation from videos requires systems to retrieve relevant audiovisual evidence from large corpora and synthesize it into coherent, attributed text. Current approaches struggle at both ends: retrieval methods fail on complex, multi-faceted queries that cannot be captured by a single embedding, while generation methods lack the high-level reasoning needed to synthesize across multiple videos and face memory constraints over long, multi-video contexts. We present \marquis: a three-stage pipeline that addresses these limitations through (1) query expansion, fusion, and reranking, (2) calibrated structured evidence extraction, and (3) article generation from extracted evidence, optionally controlled by an \rlm. On the MAGMaR2026 shared task, we improve retrieval performance from 0.195 to 0.759 (nDCG@10). For article generation, \iterqabase improves average human score from 3.09 to 3.83 over the \cag baseline, while \marquis-\rlm achieves a human score of 3.30 and the strongest citation recall among non-QA systems.\footnote{We release the code here: \url{https://github.com/debashishc/marquis}}

\end{abstract}

\section{Introduction}
\label{sec:intro}
Large-scale video corpora now document real-world events with a breadth and immediacy that no single text source can match, yet turning this raw audiovisual evidence into a well-sourced analytical article remains largely a manual process. Grounded article generation \cite{martin2025wikivideoarticlegenerationmultiple} from large video collections requires systems to retrieve relevant audiovisual evidence and synthesize it into coherent, attributable text. 

Current video retrieval and generation systems each face distinct limitations. Retrieval methods struggle with complex information needs that combine multiple implicit and explicit sub-needs and instructions in a single query \cite{weller2024followirevaluatingteachinginformation}, failing to surface all relevant videos to the information request. Generation methods face three interrelated challenges: long multi-video contexts exceed model memory constraints \cite{chen2024longvilascalinglongcontextvisual, he2024malmmmemoryaugmentedlargemultimodal, li2025videochatflashhierarchicalcompressionlongcontext}, existing VLMs are not designed for multi-video reasoning, and most video understanding work remains focused on low-level recognition tasks like captioning and entity-centric QA rather than the high-level synthesis required for article generation \cite{martin2025wikivideoarticlegenerationmultiple}. These limitations compound in a full pipeline: retrieval errors propagate missing or irrelevant evidence into generation, where models already struggle to reason over the context they receive.

In this work, we present \marquis (\textbf{M}ultimodal \textbf{A}rticle generation via \textbf{R}etrieval, \textbf{Q}uery decomposition, \textbf{U}ncertainty calibration, and \textbf{I}terative evidence \textbf{S}ynthesis), a three-stage pipeline that addresses these limitations through modular decomposition of retrieval, evidence extraction, and generation. First, we decompose and expand each query into sub-queries, retrieve independently over each sub-query, and fuse the resulting ranked lists before reranking. Second, we extract evidence from retrieved videos through complementary query-agnostic and query-conditioned components, then calibrate each extracted claim against its source video to estimate support probability. Third, we generate cited articles from the curated evidence, comparing single-prompt, clustering-based, and bullet-point strategies. We additionally introduce \marquis-\rlm, an instantiation of Recursive Language Models \citep[RLM;][]{zhang2025recursive} that treats each pipeline module as a callable tool within a persistent structured-memory environment, enabling iterative evidence gathering, cross-video conflict resolution, and fact curation before generating the final article. Our contributions can be summarized as follows:
\begin{enumerate}
    \item We introduce \marquis, a three-stage pipeline for large-scale video retrieval-augmented article generation.
    \item Our two-stage retrieval approach, combining query expansion with rank fusion and video-native reranking, improves nDCG@10 from 0.195 to 0.759 over a dense retrieval baseline on MAGMaR2026.
    \item Our QA-based article generation approach, combining query decomposition with video-grounded question answering, improves average human score from 3.09 to 3.83 over the \cag baseline on MAGMaR2026 oracle article generation.
\end{enumerate}

\section{Related Work}
\label{sec:related}
\subsection{Multimodal Retrieval and RAG}

Retrieval-augmented generation \citep[RAG;][]{lewis2021retrievalaugmentedgenerationknowledgeintensivenlp} grounds language model outputs in retrieved evidence. \citet{martin2025wikivideoarticlegenerationmultiple} formalize multi-video article generation, the task of retrieval-augmented generation from multiple videos. 

\paragraph{Retrieval}
Retrieving videos has been widely studied, but \citet{kriz2025multivent20massivemultilingual} show that most methods are specialized to descriptive queries and do not generalize to semantic queries or scale to large corpora. However, bi-encoder methods for dense \citep{luo2021clip4clipempiricalstudyclip, zhu2024languagebindextendingvideolanguagepretraining, ma2025tevatron20unifieddocument, li2026qwen3vlembeddingqwen3vlrerankerunifiedframework}, multi-vector \citep{reddy2025videocolbertcontextualizedlateinteraction, qin2026multivectorindexcompressionmodality}, and modality fusion \cite{samuel2025mmmorrfmultimodalmultilingualmodularized} provide scalable options for retrieval. Video reranking \cite{li2026qwen3vlembeddingqwen3vlrerankerunifiedframework, skow2026rankvideoreasoningrerankingtexttovideo} helps balance performance and scalability further, reranking the outputs of first-stage bi-encoder methods. 

\paragraph{Generation}
Most work generating text from videos focuses on low-level extraction and single-video settings such as captioning and question answering  \citep{msrvtt, krishna2017densecaptioningeventsvideos, lei2019tvqalocalizedcompositionalvideo, yu2019activitynetqadatasetunderstandingcomplex, zhang2025hltcoeevaluationteamtrec}. \citet{martin2025wikivideoarticlegenerationmultiple} show that existing VLMs fixate on low-level visual features and fail at the high-level synthesis required for article generation. 

\marquis{} differs by integrating retrieval, calibrated extracted claims, QA-based evidence extraction, and iterative evidence control into a single system.

\subsection{Long-Context Video Understanding}
Recent long-video models extend temporal range through long-context training \citep{li2025videochatflashhierarchicalcompressionlongcontext}, memory augmentation \citep{he2024malmmmemoryaugmentedlargemultimodal}, and hierarchical compression \citep{chen2024longvilascalinglongcontextvisual}, but still face computational and reasoning limits over extended multimodal context. Additionally, none of these methods are trained for multi-video settings.
Recursive Language Models \citep[RLMs;][]{zhang2025recursive} externalize long inputs into an interactive environment that can be inspected and processed through programmatic operations. We use this idea as a control layer for article generation: rather than placing all extracted evidence into one context, \marquis-\rlm{} iteratively gathers, stores, and curates extracted claims before writing.

\begin{figure*}
    \centering
    \includegraphics[width=\linewidth]{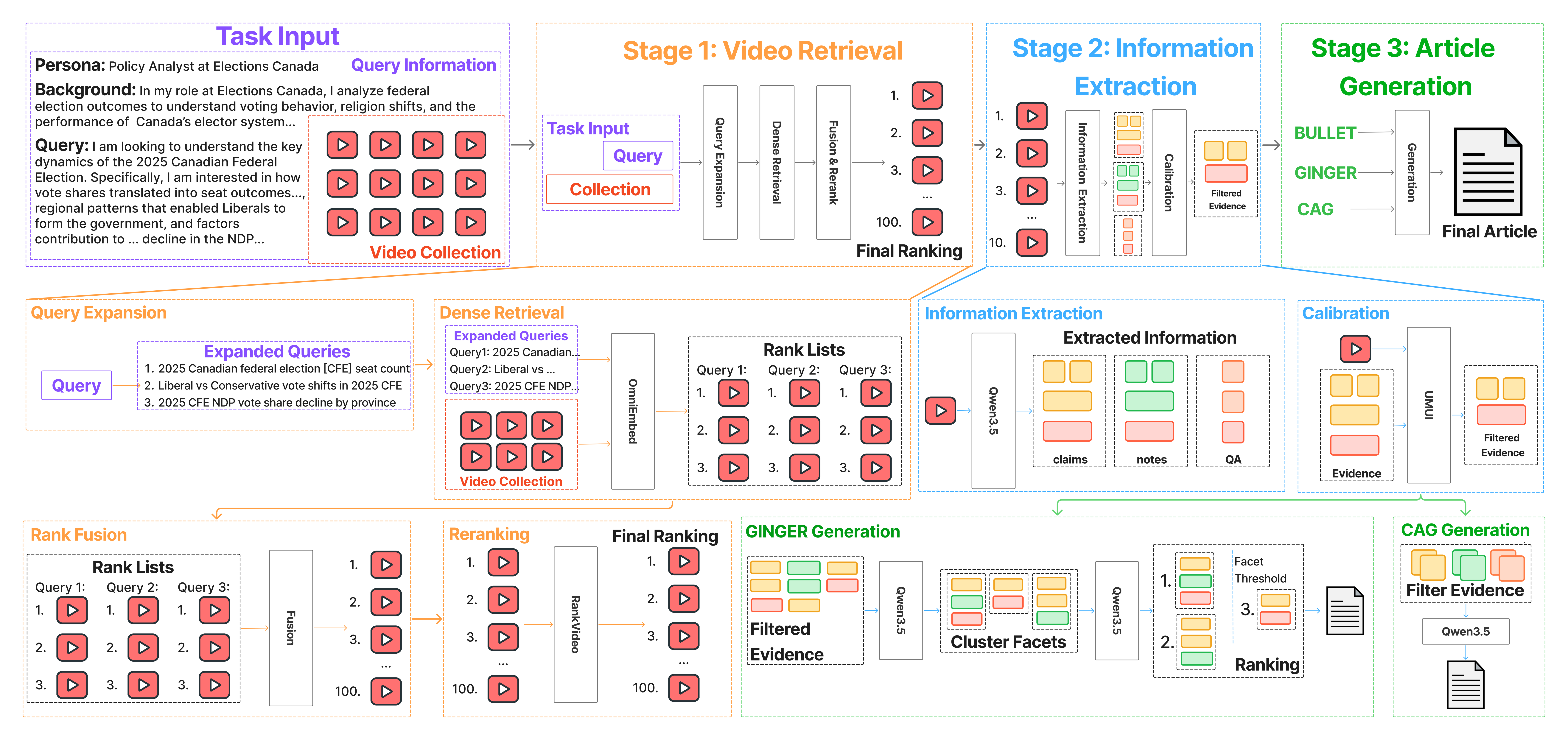}
    \caption{Overview of \marquis. \textbf{\textcolor{orange}{Stage 1 (Video Retrieval):}} Each query is decomposed into sub-queries, which are independently encoded by OmniEmbed and retrieved against the corpus. The resulting ranked lists are fused and reranked by \rankvideo to produce the final ranking. \textbf{\textcolor{CornflowerBlue}{Stage 2 (Information Extraction):}} Videos are processed by parallel information extraction streams---query-conditioned claims, query-agnostic notes, and QA---using Qwen3.5. Extracted evidence is scored by CLUE, a calibrated multimodal uncertain-inference model trained for the Unified Multimodal Uncertain Inference (UMUI) task, to filter unsupported claims. \textbf{\textcolor{Green}{Stage 3 (Article Generation):}} The filtered evidence is then passed to an article generator: \bulletmethod passes the list of extracted claims as the article, \cag single-prompt baseline, or \ginger-based generation to produce a final cited article.}
    \label{fig:marquis-overview}
\end{figure*}

\section{Retrieval}
\label{sec:retrieval}
Our retrieval pipeline operates in two stages. First, we decompose each  query into atomic sub-queries, retrieve independently over each, and fuse the resulting ranked lists into a single candidate set. Second, we rerank the fused candidates using a video-native reranker. \textcolor{red}{\autoref{fig:marquis-overview}} illustrates the full pipeline.

\subsection{First-Stage}
Our first-stage method consists of three key components: (1) Query Decomposition and Expansion, (2) Dense Retrieval, and (3) Rank Fusion.

\paragraph{Query decomposition and expansion.} We focus on queries that are long, instructional requests that combine a professional persona, domain background, and multi-faceted information need. Dense retrievers, however, are typically trained on short, single-intent queries, and encoding a complex request as a single vector collapses its many sub-needs into one point in embedding space. To bridge this gap, we decompose each query into N atomic sub-queries using an LLM, where each sub-query targets a single retrievable fact and is phrased as a concise search phrase.

\paragraph{Dense Retrieval.}
Both original queries and decomposed sub-queries are retrieved against an omnimodal index, which produces a ranked list of the top 1,000 candidates for each query.

\paragraph{Rank fusion.} 
Given $N$ ranked lists per query, we aggregate them into a single ranking. Let $\text{rank}(v, q_i)$ denote the rank of video $v$ in the list produced by sub-query $q_i$, and let $s(v, q_i)$ denote the cosine similarity score. We evaluate five fusion strategies:

\begin{itemize}
    \item \textbf{Reciprocal Rank Fusion (RRF).} Scores each video by the sum of reciprocal ranks across sub-query lists, with smoothing constant $K \in \{10, 60, 100\}$:
    \begin{equation}
        \text{RRF}_K(v) = \sum_{i=1}^{N} \frac{1}{K + \text{rank}(v, q_i)}
    \end{equation}

    \item \textbf{Sum of similarities.} Scores each video by the total cosine similarity across all sub-queries:
    \begin{equation}
        \text{Score}_{\text{sum}}(v) = \sum_{i=1}^{N} s(v, q_i)
    \end{equation}

    \item \textbf{Max similarity.} Scores each video by its highest similarity across sub-queries:
    \begin{equation}
        \text{Score}_{\text{max}}(v) = \max_{i} \; s(v, q_i)
    \end{equation}

    \item \textbf{Mean similarity.} Scores each video by the average similarity across sub-queries:
    \begin{equation}
        \text{Score}_{\text{mean}}(v) = \frac{1}{N} \sum_{i=1}^{N} s(v, q_i)
    \end{equation}

    \item \textbf{Weighted RRF.} Weights each reciprocal rank contribution by its cosine similarity:
    \begin{equation}
        \text{WRRF}_K(v) = \sum_{i=1}^{N} \frac{s(v, q_i)}{K + \text{rank}(v, q_i)}
    \end{equation}
\end{itemize}

Implementation details and expansion statistics are provided in \autoref{app:retrieval}.

\subsection{Reranking}

Given the top 100 videos per query from the first-stage retrieval, we perform reranking with RankVideo \cite{skow2026rankvideoreasoningrerankingtexttovideo}. For each full query, we pass the candidate videos from the first-stage retrieval to RankVideo and reorder the ranked list based on these judgments.

\section{Information Extraction}
\label{sec:info-extraction}
The retrieval stage returns video candidates, but article generation requires finer-grained evidence. We therefore convert retrieved videos into extracted claims that can be selected, calibrated, cited, and passed to the generation stage. Our evidence extraction system contains three components: \textit{query-agnostic} note extraction, \textit{query-conditioned} claim extraction, and question-answer extraction. The components differ in how they condition on the query, but all produce source-linked extracted evidence with video identifiers and, when available, timestamps. The extraction component is shown at a high level in \autoref{fig:marquis-overview}; implementation details are provided in \autoref{fig:info-extraction}.

Let $V(q)$ denote the videos associated with query $q$. For each video $v \in V(q)$, the extraction stage may produce three evidence families:
\[
N(v) = \{n_1, \dots, n_{|N(v)|}\},
\]
\[
 C(v,q) = \{c_1, \dots, c_{|C(v,q)|}\},
\]
\[
A(v,q) = \{a_1,\ldots,a_{|A(v,q)|}\},
\]
where $N(v)$ denotes query-agnostic notes, $C(v,q)$ denotes query-conditioned claims, and $A(v,q)$ denotes question-answer outputs. Each output is later scored against its source video by the shared calibration stage.

\subsection{Query-Agnostic Note Extraction}
\label{sec:info-ext-general-notes}
The query-agnostic component builds a reusable evidence base from each video without conditioning on a specific information need. Its goal is to capture directly observable visual events, on-screen text, and spoken content that may be useful across queries. The extractor is prompted to avoid speculation, causal inference, and cross-video synthesis. Each note describes a single atomic observation and includes a modality tag and optional timestamp. Confidence is not assigned at extraction time; support is estimated in a separate post-extraction calibration stage, which avoids conflating evidence extraction with support estimation.

\subsection{Query-Conditioned Claim Extraction}
\label{sec:info-ext-query-claims}
The query-conditioned component targets evidence extraction toward a specific query. Where general notes prioritize breadth, this component prioritizes task relevance: given a specific  information need, it extracts only claims that are relevant to that need. For each query-video pair, the extractor receives the query, persona, background, topic, and video metadata, and returns claims that are both query-relevant and directly supported by the video. This stage is not free-form answer generation: the prompt explicitly discourages generic scene descriptions, unsupported inferences, and redundant paraphrases. Each claim record contains a claim identifier, query identifier, video identifier, topic label, claim text, and optional support-oriented fields such as confidence, evidence description, source type, and timestamp. The resulting claims provide a targeted evidence set for downstream article generation.

\subsection{Question-Answer Extraction}
\label{sec:info-ext-qa}
The question-answering component extracts evidence through targeted video question answering for information needs. Given a query, persona, and background, the system decomposes the information need into atomic subqueries, retrieves relevant videos for each subquery, and uses a vision-language model to answer using the retrieved video content and transcript. We implement two variants: a single-shot variant that answers a fixed set of decomposed subqueries and aggregates the grounded per-video answers without introducing external knowledge, and an iterative variant that generates follow-up questions conditioned on previous question-answer history to pursue missing or underspecified information. The output is a set of question-answer evidence records, each linked to the question, answer, source video, and any available timestamp or confidence metadata.

\subsection{Video-Grounded Calibration}
\label{sec:info-ext-calibration}
After extraction, each output is scored against its source video. Given a video $v$ and artifact $x \in N(v) \cup C(v,q) \cup A(v,q)$, calibration produces a support probability
\[
s_\theta(v,x) \in [0,1].
\]
The score estimates whether the output is supported by the source video. Calibration is applied after extraction so that evidence creation and support estimation remain separate. The calibrated outputs retain their original text and metadata, with the support score attached for downstream filtering and article generation. Implementation details and prompts are provided in \autoref{app:calibration} and \autoref{app:prompts}.

\section{Article Generation}
\label{sec:report_generation}
Our article generation pipeline synthesizes extracted video evidence into a fluent, source-attributed article that answers the information need. This stage operates after retrieval, evidence extraction, QA, and calibration. The article generator does not inspect raw videos directly, but instead consumes structured evidence artifacts tied to source videos and, when available, timestamps.

We design the article generation stage to be input-agnostic. In the experiments reported here, the same generation procedures can operate over query-conditioned claims, query-agnostic notes, or question--answer pairs produced by the QA pipeline. Claims and notes include explicit video identifiers and timestamps, while QA pairs include the source videos used to produce the answer. The generator receives a flat list of extracted evidence together with their source metadata and is instructed to produce an article whose factual statements are supported by inline citations.

We compare three article generation strategies.

\paragraph{\bulletmethod{}.}
The simplest strategy does not synthesize evidence into prose. It renders the retrieved evidence items directly as a numbered list of findings with inline citations. This output is conservative and preserves the connection between evidence and source videos, but it does not produce the coherent article-style response required by the task.

\paragraph{\cag} is our Collaborative Article Generation baseline, adapted from WikiVideo \cite{martin2025wikivideoarticlegenerationmultiple} to operate over extracted evidence. It generates a cited article from the extracted evidence for a query in one synthesis pass, following the role of the text-only aggregator in WikiVideo \cag{}. The model is instructed to organize evidence, remove redundancy, and preserve citations.

\paragraph{\ginger{}.}

We adapt the \ginger{} framework \cite{ajewska2025GINGERGI} to video-grounded extracted evidence. Since our extraction stage already produces atomic evidence, we skip nugget detection and perform facet clustering, cluster ranking, per-cluster summarization, and fluency enhancement. The model first groups evidence into thematic clusters (e.g., casualties, rescue efforts, government response), ranks them by query relevance, summarizes the top clusters independently into short cited sentences, and finally rewrites them into a coherent article. This staged-decision decomposes article generation into smaller controlled calls and helps preserve citations.

\section{\rlm{} Controller}
\label{sec:rlm}

In addition to the aforementioned pipeline, we further construct a \rlm{}-based high-level control system for article generation, \marquis-\rlm{}, organizing each module of the pipeline as a tool to be called by the \rootlm{}. 
 
Conceptually, \rlm{} serves here as a general recursive control paradigm, whereas \marquis-\rlm{} is our task-specific instantiation of this paradigm for multi-video article generation.
Unlike standard code-generating \rlm{}, \marquis-\rlm{} equips \rootlm{} to call our pre-developed modules in REPL, preserving robust performance of specialized modules while gaining the reasoning and efficiency of the \rlm{} paradigm.
We further make explicit structured memory a core component of the system: \rootlm{} always reasons over an evidence record that can be searched, reused, and revised, rather than relying only on what remains in context. This mitigates the information forgetting and multi-source confusion common in long iterative workflows, while making cross-video conflicts and missing information explicit. Examples are provided in \autoref{app:rlm}.

\paragraph{REPL Environment and Tool Interface.}
\label{sec:rlm-repl-tool}
We instantiate \marquis-\rlm{} in a persistent Python sandbox whose namespace contains task context, memory bank, and callable sub-tools adapted from the modules in previous sections. At each iteration, the \rootlm{} generates and executes code in the persistent namespace, accesses raw video, audio, and transcripts only through callable tools.

\paragraph{Memory Bank.}
\label{sec:rlm-memory}

Under MAGMaR’s long-context setting, the \rootlm{} could face recurring failures including evidence forgetting, cross-source confusion, and missed conflicts or information gaps.
All stem from the same limitation: the \rootlm{} can only reason over what remains visible in context \cite{shi2025look,zhang2025memory}. Even the original free-form RLM-style REPL state is insufficient, since it introduces naming drift, reassignment errors, schema drift, and perception-level confusion.
To address this, inspired by recent external-memory designs for LM agents such as MemR$^3$ \cite{du2025memr3memoryretrievalreflective}, we build a dynamic structured memory on top of the REPL. 

The full schema and operators are given in \autoref{app:rlm}.

\paragraph{Think--Act--Observe.}
\label{sec:rlm-tao}
We require the \rootlm{} to follow a coarse-grained Think–Act–Observe mechanism, inspired by interleaved reasoning-and-acting frameworks such as ReAct~\cite{yao2022react}. 

This design enforces immediate external feedback at each step and grounds reasoning in explicit state transitions, while still leaving tool choice and reflection frequency to the \rootlm{}.

\section{Experiments}
\label{sec:results}
\newcommand{\green}[1]{\textcolor{green!50!black}{#1}}
\newcommand{\red}[1]{\textcolor{red!70!black}{#1}}

\begin{table*}
  \centering
  \begin{tabular}{ll| ccc ccc}
    \toprule
    & \textbf{Run} & \textbf{nDCG@10} & \textbf{nDCG@20} & \textbf{nDCG@100} & \textbf{R@10} & \textbf{R@20} & \textbf{R@100} \\
    \midrule
    & OmniEmbed & 0.195 & 0.229 & 0.311 & 0.190 & 0.276 & 0.494 \\
    \midrule
    \multirow{3}{*}{\textit{Sim}}
    & Max Sim & \textbf{0.722} & \textbf{0.743} & \textbf{0.784} & \textbf{0.639} & 0.731 & 0.826 \\
    & Mean Sim & 0.637 & 0.650 & 0.696 & 0.544 & 0.618 & 0.736 \\
    & Sum Sim & 0.703 & 0.725 & 0.776 & 0.604 & 0.698 & 0.818 \\
    \midrule
    \multirow{4}{*}{\textit{RRF}}
    & K=10 & 0.700 & 0.739 & 0.777 & 0.612 & \textbf{0.735} & \textbf{0.832} \\
    & K=60 & 0.695 & 0.728 & 0.773 & 0.599 & 0.714 & 0.823 \\
    & K=100 & 0.688 & 0.719 & 0.767 & 0.590 & 0.704 & 0.818 \\
    & Weighted & 0.699 & 0.730 & 0.778 & 0.604 & 0.714 & \textbf{0.832} \\
    \bottomrule
  \end{tabular}
  \caption{First-stage retrieval results. Best score per column is \textbf{bolded}.}
  \label{tab:first-stage-retrieval}
\end{table*}

\paragraph{Dataset}
We evaluate on the MAGMaR2026 Test Set. The data is based on WikiVideo \cite{martin2025wikivideoarticlegenerationmultiple}. For the retrieval and RAG settings, we retrieve relevant videos from a combination of MAGMaR and MultiVENT2.0 \cite{kriz2025multivent20massivemultilingual}. For oracle article generation, systems receive the ground-truth relevant videos, isolating generation quality from retrieval quality.

\paragraph{Evaluation Setup}
Retrieval results are evaluated with nDCG and Recall for 10, 20, and 100. We use the ir-measures \cite{ir-measures} to calculate these scores. Generated articles are evaluated with an automatic and human evaluation. For the automatic evaluation, the systems are evaluated by MiRAGE \cite{martin2025seeingmirageevaluatingmultimodal}, which captures the factuality, information coverage, groundedness, and proper attribution of citations. Each MiRAGE entailment judgment is judged by CLUE \cite{zhang2026unified}. For human evaluation, three human annotators provide scalar scores of 1--5 for each system, scoring factuality, adequacy, coherence, relevancy, and fluency. After providing scalar scores for each prediction, the annotators also pick the best system response to each query.

\paragraph{Experimental Setup.}
All systems are evaluated under the same MAGMaR2026 retrieval and oracle-generation splits, but differ in their video access patterns and model backends. Retrieval uses OmniEmbed \cite{ma2025tevatron20unifieddocument} for video and query encoding and Qwen3.5-9B \cite{qwen35} for query decomposition. The information-extraction streams (note and claim extraction, calibration) use Qwen3.5-9B over sampled video frames. The QA pipeline combines Qwen3.5-27B for answer generation, Qwen2.5-Omni-7B \cite{xu2025qwen25omnitechnicalreport} with OmniEmbed for multimodal embeddings, and Whisper medium.en \cite{Radford2022-xl} for transcription. Article generation uses Qwen3.5-27B for both the single-prompt baseline and \ginger{}-based generators. The \rlm{} controller runs a Qwen3.5-9B root LM that calls the extraction and QA modules as sub-tools. None of the claim-based extraction or generation systems use audio; only the QA pipeline and \rlm{} (via its transcription tool) access the audio stream. Frame rates, top-$k$ values, generation budgets, and other component-specific hyperparameters are listed in \autoref{app:setup}.

\begin{table*}
  \centering
  \setlength{\tabcolsep}{4pt}
  \begin{tabular}{ll| ccc ccc}
    \toprule
    & \textbf{Run} & \textbf{nDCG@10} & \textbf{nDCG@20} & \textbf{nDCG@100} & \textbf{R@10} & \textbf{R@20} & \textbf{R@100} \\
    \midrule
    & OmniEmbed + RV & 0.542 & 0.534 & 0.546 & 0.423 & 0.462 & 0.494 \\
    & & \green{177.95} & \green{133.19} & \green{75.56} & \green{122.63} & \green{67.39} & N/A \\
    \midrule
    \multirow{6}{*}{\textit{Sim}}
    & Max Sim + RV & 0.399 & 0.405 & 0.425 & 0.344 & 0.383 & 0.437 \\
    & & \red{-44.74} & \red{-45.49} & \red{-45.79} & \red{-46.17} & \red{-47.61} & \red{-47.09} \\
    & Mean Sim + RV & 0.740 & 0.723 & 0.750 & 0.637 & 0.665 & 0.736 \\
    & & \green{16.17} & \green{11.23} & \green{7.76} & \green{17.10} & \green{7.61} & N/A \\
    & Sum Sim + RV 
        & 0.747 & 0.758 & 0.800 
        & 0.636 & 0.711 & 0.818 \\
    & 
        & \green{6.26} & \green{4.55} & \green{3.09} 
        & \green{5.30} & \green{1.86} & N/A \\
    \midrule
    \multirow{8}{*}{\textit{RRF}}
    & RRF K=10 + RV 
        & \textbf{0.759} & \textbf{0.771} & \textbf{0.811} 
        & \textbf{0.652} & \textbf{0.735} & \textbf{0.832} \\
    & 
        & \green{8.43} & \green{4.33} & \green{4.38} 
        & \green{6.54} & N/A & N/A \\
    & RRF K=60 + RV 
        & 0.754 & 0.765 & 0.807 
        & 0.641 & 0.716 & 0.823 \\
    & 
        & \green{8.49} & \green{5.08} & \green{4.40} 
        & \green{7.01} & \green{0.28} & N/A \\
    & RRF K=100 + RV 
        & 0.746 & 0.757 & 0.799 
        & 0.636 & 0.711 & 0.818 \\
    & 
        & \green{8.43} & \green{5.29} & \green{4.17} 
        & \green{7.80} & \green{0.99} & N/A \\
    & Weighted RRF + RV 
        & 0.757 & 0.768 & 0.810 
        & 0.650 & 0.725 & \textbf{0.832} \\
    & 
        & \green{8.30} & \green{5.21} & \green{4.11} 
        & \green{7.62} & \green{1.54} & N/A \\
    \bottomrule
  \end{tabular}
  \caption{Reranked retrieval results with percentage change relative to first-stage baseline. \green{Green} denotes improvement, \red{red} denotes degradation.}
  \label{tab:reranked-retrieval}
\end{table*}
\subsection{Retrieval}



In \autoref{tab:first-stage-retrieval} and \autoref{tab:reranked-retrieval}, we report the results of video retrieval for first-stage and reranking, respectively. All query expansion and fusion methods substantially outperform the OmniEmbed dense retrieval baseline. This confirms that decomposing complex queries into sub-queries targeting atomic pieces of information is much more suitable for a dense retriever. This is an intuitive result, as most first-stage retrievers are trained on short, single-intent query-document pairs and compressing a complex information request into a single embedding is out-of-distribution and challenging  \cite{weller2024followirevaluatingteachinginformation}. However, our sub-queries reduce this burden, allowing for the model to interface with in-distribution queries and leaving the merging of those ranked lists to a fusion or reranking approach.

\paragraph{Similarity vs.\ RRF fusion.}
Among first-stage methods (\autoref{tab:first-stage-retrieval}), Max similarity achieves the highest nDCG at all cutoffs. It benefits from its selection mechanism: because it scores each video by its best-matching sub-query, it surfaces videos that are highly relevant to at least one facet of the information need, even if they are irrelevant to others. RRF strategies \cite{cormack-rrf}, which aggregate evidence across all sub-queries, produce more balanced rankings and achieve higher recall at deeper cutoffs, suggesting they are better at capturing the full breadth of a multi-faceted query. Mean and Sum similarity consistently underperforms the other aggregation methods, likely because averaging dilutes strong matches with weak ones. Among the RRF variants, lower $K$ values perform slightly better, as a smaller smoothing constant amplifies rank differences and rewards videos that appear near the top of multiple sub-query lists. Weighted RRF performs comparably to standard RRF, indicating that weighting reciprocal ranks by cosine similarity provides limited additional signal when the sub-queries are already well-targeted.

\paragraph{Reranking.}
As shown in \autoref{tab:reranked-retrieval}, applying RankVideo reranking improves performance across nearly all fusion strategies. Among the expanded-query methods, all RRF variants and similarity variants (except Max) see consistent improvements, with RRF at $K{=}10$ achieving the best ranking performance overall. The one notable exception is Max similarity, where reranking sharply degrades all metrics. We leave a detailed analysis of this failure mode to future work.

\begin{table*}[t]
\centering
\begin{tabular}{l|ccccccc}
\toprule
\multirow{2}{*}{System} & \multirow{2}{*}{Human Score} & \multirow{2}{*}{Best Votes} & \multirow{2}{*}{Best \%} & \multicolumn{2}{c}{Info} & \multicolumn{2}{c}{Cite} \\
\cmidrule(lr){5-6} \cmidrule(lr){7-8}
& & & & P & R & P & R \\
\midrule
\cag{} (baseline) & 3.09 & 1 & 1.8\% & \underline{76.4} & \textbf{41.0} & \underline{61.7} & 22.8 \\
\midrule
\bulletmethod{} & 2.67 & 0 & 0.0\% & 71.1 & 39.4 & 60.4 & 23.7 \\
\midrule
\ginger{} & 3.12 & 6 & 10.5\% & \textbf{77.6} & \underline{40.4} & \textbf{64.3} & 22.6 \\
\marquis-\rlm{} & 3.30 & 3 & 5.3\% & 70.8 & 38.5 & 59.2 & \underline{27.2} \\
\midrule
SS QA Base & 3.07 & 6 & 10.5\% & 33.1 & 30.6 & 27.7 & \textbf{28.1} \\
\ssqaginger{} & 3.42 & \textbf{10} & \textbf{17.5\%} & 54.4 & 32.4 & 32.6 & 23.8 \\
\midrule
\iterqabase{} & \textbf{3.83} & \underline{8} & \underline{14.0\%} & 34.7 & 31.3 & 26.8 & 25.8 \\
\iterqaginger{} & \underline{3.69} & 5 & 8.8\% & 34.5 & 29.0 & 25.7 & 22.6 \\
\bottomrule
\end{tabular}
\caption{Oracle generation results for \marquis{} systems. H = human score; B = best-system votes; IP/IR = information precision/recall; CP/CR = citation precision/recall.}
\label{tab:oracle-generation-mars}
\end{table*}
\subsection{Generation}

In \autoref{tab:oracle-generation-mars}, we report oracle generation results, where each system receives the ground-truth relevant videos rather than retrieved candidates, isolating generation quality from retrieval effects. We evaluate eight system variants spanning three evidence pipelines: claim-based extraction (\bulletmethod{}, Ginger), question answering (iterative (Iter QA) and single shot (SS QA)), and RLM-controlled generation.

\paragraph{Generation Systems.}
Among the claim-based generation variants, \ginger{} is the strongest prose generator, improving over the \cag{} baseline in human score, best-vote share, and both information and citation precision. Its staged decomposition into facet clustering, ranking, and per-cluster summarization appears to help the model organize evidence and preserve citations more reliably than a single generation call. \bulletmethod{} shows the opposite tradeoff: it achieves slightly higher citation recall than the other claim-based systems, but receives the lowest human score and no best-system votes, confirming that annotators penalize outputs that lack fluent synthesis even when source attribution is preserved. Taken together, these results suggest that explicit topical organization improves generation quality, but that the final output must still read as coherent prose to satisfy analyst information needs.

\paragraph{QA Systems.}
QA-based systems achieve the strongest human preference scores. \iterqabase{} obtains the highest average human score, while \ssqaginger{} receives the most best-system votes. Their aggregate automatic metrics are weaker, largely because QA failures on a small number of topics produce conservative empty or near-empty outputs. This suggests that QA improves article usefulness when relevant answers are recovered, but remains brittle when decomposition or video-level answering fails. When the QA systems fail to answer questions, due to VLM failures or irrelevant sub-questions, the downstream generation systems often refuse to write the article, backing off due to insufficient evidence. This conservative behavior is a double-edged sword: it avoids hallucination on topics where the video evidence genuinely lacks the requested information (e.g., Myanmar Earthquake Q1), but it produces zero-score outputs that sharply deflate aggregate metrics on topics where information was available.

\paragraph{\rlm{}.}
\marquis-\rlm{} improves human score over \cag{} and \bulletmethod{} and achieves the highest citation recall among non-QA systems. This suggests that iterative evidence gathering and structured memory help preserve attribution across multi-video contexts. Its lower precision and citation precision, however, indicates that the controller also admits less relevant facts into the final article. We therefore view \marquis-\rlm{} as an evidence-management mechanism rather than a standalone replacement for structured generation; its Think--Act--Observe loop and persistent memory bank are effective at resolving cross-video conflicts and filling information gaps (see examples in \autoref{app:rlm}), and tighter integration with \ginger{}-based synthesis is a natural next step.

\section{Conclusion}
\label{sec:Conclusion}
We presented \marquis, a three-stage pipeline for video retrieval-augmented article generation. \marquis{} decomposes complex queries, retrieves and reranks relevant videos, converts video content into calibrated extracted evidence, and generates cited articles from selected evidence. The optional \marquis-\rlm{} controller extends this pipeline by treating retrieval, extraction, QA, calibration, and generation as tools within a structured-memory environment, enabling iterative evidence gathering and curation before writing.
Our experiments show that explicit query decomposition and video-native reranking substantially improve retrieval, while article-generation results reveal complementary tradeoffs among claim-based, QA-based, and \rlm-controlled systems. More broadly, our findings suggest that grounded generation from video is best framed as an evidence-management problem. Rather than prompting a model to summarize long multi-video context directly, effective systems should retrieve broadly, extract atomically, estimate source support, and synthesize only from selected extracted evidence. Future work should improve learned calibration, integrate retrieval and extraction more tightly, and develop generation methods that combine structured evidence organization with iterative evidence control.

\section*{Acknowledgment}
This material is based upon work supported by the NSF Graduate Research Fellowship under Grant No. DGE2139757. Any opinion, findings, and conclusions or recommendations expressed in this material are those of the author(s) and do not necessarily reflect the views of the National Science Foundation.

\bibliography{custom}

\appendix

\newtcolorbox{promptbox}[1][]{
  breakable,
  enhanced,
  colback=gray!4,
  colframe=black!35,
  boxrule=0.5pt,
  arc=1mm,
  left=6pt,
  right=6pt,
  top=6pt,
  bottom=6pt,
  fontupper=\footnotesize\ttfamily,
  #1
}

\newtcolorbox{examplebox}[1][]{
  breakable,
  enhanced,
  colback=black!1,
  colframe=black!25,
  boxrule=0.5pt,
  arc=1mm,
  left=6pt,
  right=6pt,
  top=6pt,
  bottom=6pt,
  fontupper=\footnotesize\ttfamily,
  #1
}

\newcommand{\pline}[1]{#1\par}

\section{Experimental Setup Details}
\label{app:setup}

This appendix summarizes the model backends, input access, and hyperparameters used by each component. We report model backends and hyperparameters for retrieval, information extraction, calibration, and QA in \autoref{tab:unified-hparams}, and article generation hyperparameters in \autoref{tab:article-hparams}. Prompt templates are listed in \autoref{app:prompts}.
\begin{table*}[t]
\centering
\setlength{\tabcolsep}{3.5pt}
\small
\begin{tabular}{llll}
\toprule
\textbf{Component} & \textbf{Backend} & \textbf{Setting} & \textbf{Value} \\
\midrule
\multirow{8}{*}{Retrieval}
  & \multirow{8}{*}{\shortstack[l]{OmniEmbed;\\RankVideo}}
  & Decomposition model & Qwen3.5-27B ($T{=}0.7$, top-$p{=}0.9$, 2048 tok) \\
  & & Thinking mode & Disabled \\
  & & Embedding pooling / norm & End-of-sequence; L2 \\
  & & Precision & bfloat16 \\
  & & Corpus size & 109{,}814 videos \\
  & & First-stage depth & 100 videos per (sub-)query \\
  & & Fusion methods & Max, Mean, Sum, RRF, WRRF ($K{\in}\{10,60,100\}$) \\
  & & Reranking depth & 100 videos \\
\midrule
\multirow{5}{*}{\shortstack[l]{Note / Claim\\Extraction}}
  & \multirow{5}{*}{Qwen3.5-9B}
  & FPS / max frames & 1.0 / 128 \\
  & & Decoding & $T{=}0.3$, top-$p{=}0.8$, top-$k{=}20$ \\
  & & Max tokens (notes / claims) & 2048 / 4096 \\
  & & Seed (notes / claims) & 42 / 40 \\
  & & Thinking & off \\
\midrule
\multirow{3}{*}{Calibration}
  & \multirow{2}{*}{\shortstack[l]{CLUE;\\prompted Qwen3.5}}
  & FPS & 0.5 \\
  & & Frame size & $256 \times 256$ \\
  & & Filtering threshold & 0.5 \\
\midrule
\multirow{8}{*}{QA}
  & \multirow{6}{*}{\shortstack[l]{Qwen3.5-27B;\\OmniEmbed;\\Whisper medium.en}}
  & Questions / query & 10--25 \\
  & & Question decoding & $T{=}0.4$, top-$p{=}0.9$, 1024 tok \\
  & & Video QA / aggregation tokens & 512 / 256 \\
  & & Iterative max steps & 5 / question \\
  & & Frame budget & 32 frames \\
  & & Audio & 16\,kHz mono \\
\midrule

\multirow{5}{*}{RLM}
  & \multirow{5}{*}{\shortstack[l]{Qwen3.5-9B (root, VLM);\\Qwen3.5-27B (judge);\\ tools as above}}
  & Root LM context              & 32{,}768 tokens \\
  & & Caption VLM                 & 32 frames, 32{,}000 tok \\
  & & Caption VLM decoding        & $T{=}0.3$ \\
  & & Max iterations              & 60 \\
  & & LLM-as-a-Judge              & $T{=}0.2$, 512 tok, per-iteration \\
\bottomrule
\end{tabular}
\caption{Unified component backends and hyperparameters for all \marquis{} pipeline stages.}
\label{tab:unified-hparams}
\end{table*}

\begin{table*}[t]
\centering
\setlength{\tabcolsep}{4pt}
\small
\begin{tabular}{lccc}
\toprule
\textbf{Method} & \textbf{Input} & \textbf{Max Tokens} & \textbf{Decoding} \\
\midrule
\bulletmethod{}    & Selected claims & --   & -- \\
\ginger{} cluster  & Claims          & 2048 & $T{=}.3$, $p{=}.9$ \\
\ginger{} rank     & Clusters        & 512  & $T{=}.3$, $p{=}.9$ \\
\ginger{} summarize & Top 5 clusters & 256  & $T{=}.5$, $p{=}.9$ \\
\ginger{} fluency  & Summaries       & 1024 & $T{=}.7$, $p{=}.9$ \\
\bottomrule
\end{tabular}
\caption{Article generation hyperparameters. All methods use Qwen3.5-27B.}
\label{tab:article-hparams}
\end{table*}

\section{Additional Results}
\label{append:results}
\begin{table*}[t]
\centering
\begin{minipage}{0.48\textwidth}
\centering

\begin{tabular}{ll|cc|cc}
\toprule
 & System & \multicolumn{2}{c|}{Info F1} & \multicolumn{2}{c}{Cite F1} \\
\cmidrule(lr){3-4} \cmidrule(lr){5-6}
 & & P & R & P & R \\
\midrule
\multirow{8}{*}{Q1}
    & CAG & \textbf{94.1} & \textbf{33.3} & 2.9 & 0.0 \\
\cmidrule{2-6}
    & Bullet & 77.3 & 28.6 & \textbf{9.1} & 0.0 \\
    & Ginger & 91.4 & \textbf{33.3} & 8.6 & 0.0 \\
    & RLM & 81.5 & 28.6 & 3.7 & 0.0 \\
    & Iter-B & 0.0 & 0.0 & 0.0 & 0.0 \\
    & Iter-G & 0.0 & 0.0 & 0.0 & 0.0 \\
    & SS-B & 19.4 & 3.2 & 0.0 & 0.0 \\
    & SS-G & 27.3 & 4.8 & 2.3 & 0.0 \\
\midrule
\multirow{8}{*}{Q2}
    & CAG & 87.5 & \textbf{41.3} & 4.2 & 0.0 \\
\cmidrule{2-6}
    & Bullet & \textbf{94.7} & 34.9 & 0.0 & 0.0 \\
    & Ginger & 87.5 & \textbf{41.3} & 0.0 & 0.0 \\
    & RLM & 71.4 & 38.1 & 3.6 & 0.0 \\
    & Iter-B & 0.0 & 0.0 & 0.0 & 0.0 \\
    & Iter-G & 0.0 & 0.0 & 0.0 & 0.0 \\
    & SS-B & 8.8 & 6.3 & 6.1 & 0.0 \\
    & SS-G & 35.3 & 3.2 & \textbf{12.0} & 0.0 \\
\bottomrule
\end{tabular}
\subcaption{2025 Alaskan Typhoon}
\label{tab:CLUE-2025-alaskan-typhoon}
\end{minipage}
\hfill
\begin{minipage}{0.48\textwidth}
\centering

\begin{tabular}{ll|cc|cc}
\toprule
 & System & \multicolumn{2}{c|}{Info F1} & \multicolumn{2}{c}{Cite F1} \\
\cmidrule(lr){3-4} \cmidrule(lr){5-6}
 & & P & R & P & R \\
\midrule
\multirow{8}{*}{Q1}
    & CAG & \textbf{57.1} & 36.1 & 33.3 & 13.9 \\
\cmidrule{2-6}
    & Bullet & 18.5 & 41.7 & 18.5 & 27.8 \\
    & Ginger & 50.0 & 36.1 & 28.1 & 13.9 \\
    & RLM & 27.3 & 44.4 & 21.2 & \textbf{50.0} \\
    & Iter-B & 37.0 & \textbf{52.8} & \textbf{44.4} & 47.2 \\
    & Iter-G & 22.2 & 44.4 & 27.8 & 47.2 \\
    & SS-B & 31.2 & 38.9 & 28.6 & 38.9 \\
    & SS-G & 38.5 & 33.3 & 38.5 & 38.9 \\
\midrule
\multirow{8}{*}{Q2}
    & CAG & 43.4 & 61.1 & 42.3 & 33.3 \\
\cmidrule{2-6}
    & Bullet & 46.7 & 41.7 & 46.7 & 13.9 \\
    & Ginger & 43.6 & 61.1 & 43.6 & 33.3 \\
    & RLM & \textbf{64.7} & 58.3 & \textbf{70.6} & 52.8 \\
    & Iter-B & 43.4 & 61.1 & 43.4 & 58.3 \\
    & Iter-G & 39.6 & 44.4 & 38.5 & 38.9 \\
    & SS-B & 55.0 & 69.4 & 48.3 & \textbf{69.4} \\
    & SS-G & 58.7 & \textbf{75.0} & 58.7 & 63.9 \\
\bottomrule
\end{tabular}
\subcaption{2025 Canadian Federal Election}
\label{tab:CLUE-2025-canadian-federal-election}
\end{minipage}
\caption{Per-query scores across all systems, judge: CLUE. Iter-B/G = Iter-QA-Base/Ginger, SS-B/G = SS-QA-Base/Ginger.}
\label{tab:CLUE-pair-1}
\end{table*}

\begin{table*}[]
\centering
\begin{minipage}{0.48\textwidth}
\centering

\begin{tabular}{ll|cc|cc}
\toprule
 & System & \multicolumn{2}{c|}{Info F1} & \multicolumn{2}{c}{Cite F1} \\
\cmidrule(lr){3-4} \cmidrule(lr){5-6}
 & & P & R & P & R \\
\midrule
\multirow{8}{*}{Q1}
    & CAG & 60.5 & \textbf{93.3} & 55.8 & 80.0 \\
\cmidrule{2-6}
    & Bullet & 69.8 & \textbf{93.3} & 40.0 & 0.0 \\
    & Ginger & \textbf{82.9} & 86.7 & \textbf{74.3} & \textbf{86.7} \\
    & RLM & 68.4 & 60.0 & 54.1 & 46.7 \\
    & Iter-B & 45.2 & 60.0 & 40.5 & 33.3 \\
    & Iter-G & 38.3 & 60.0 & 27.7 & 46.7 \\
    & SS-B & 17.6 & 40.0 & 15.7 & 40.0 \\
    & SS-G & 30.2 & 53.3 & 14.0 & 60.0 \\
\midrule
\multirow{8}{*}{Q2}
    & CAG & \textbf{78.1} & \textbf{93.3} & 63.3 & 80.0 \\
\cmidrule{2-6}
    & Bullet & 61.9 & 86.7 & \textbf{71.4} & 80.0 \\
    & Ginger & 76.5 & \textbf{93.3} & 60.6 & 80.0 \\
    & RLM & 73.3 & 80.0 & 63.3 & \textbf{86.7} \\
    & Iter-B & 68.4 & 86.7 & 59.6 & 66.7 \\
    & Iter-G & 54.3 & 80.0 & 48.9 & 73.3 \\
    & SS-B & 61.8 & 80.0 & 44.4 & 80.0 \\
    & SS-G & 68.6 & 86.7 & 61.2 & 73.3 \\
\bottomrule
\end{tabular}
\subcaption{2025 Myanmar Earthquake}
\label{tab:CLUE-2025-myanmar-earthquake}
\end{minipage}
\hfill
\begin{minipage}{0.48\textwidth}
\centering

\begin{tabular}{ll|cc|cc}
\toprule
 & System & \multicolumn{2}{c|}{Info F1} & \multicolumn{2}{c}{Cite F1} \\
\cmidrule(lr){3-4} \cmidrule(lr){5-6}
 & & P & R & P & R \\
\midrule
\multirow{8}{*}{Q1}
    & CAG & 88.6 & 42.9 & 71.4 & 21.4 \\
\cmidrule{2-6}
    & Bullet & \textbf{95.5} & 39.3 & 77.3 & 35.7 \\
    & Ginger & 88.2 & 42.9 & \textbf{79.4} & 21.4 \\
    & RLM & 63.0 & 42.9 & 55.6 & 42.9 \\
    & Iter-B & 48.4 & 64.3 & 43.5 & \textbf{67.9} \\
    & Iter-G & 63.5 & \textbf{71.4} & 57.7 & 46.4 \\
    & SS-B & 62.9 & 67.9 & 44.9 & 64.3 \\
    & SS-G & 72.1 & 67.9 & 63.4 & \textbf{67.9} \\
\midrule
\multirow{8}{*}{Q2}
    & CAG & 77.4 & 57.1 & 67.7 & 17.9 \\
\cmidrule{2-6}
    & Bullet & 72.2 & 57.1 & 55.6 & 46.4 \\
    & Ginger & \textbf{80.8} & 57.1 & \textbf{69.2} & 17.9 \\
    & RLM & 66.7 & 35.7 & 66.7 & 35.7 \\
    & Iter-B & 30.3 & 71.4 & 22.7 & \textbf{71.4} \\
    & Iter-G & 42.2 & 64.3 & 31.1 & 53.6 \\
    & SS-B & 29.9 & \textbf{75.0} & 25.4 & 67.9 \\
    & SS-G & 47.5 & 57.1 & 35.0 & 28.6 \\
\bottomrule
\end{tabular}
\subcaption{Blue Ghost Mission 1}
\label{tab:CLUE-blue-ghost-mission-1}
\end{minipage}
\caption{Per-query scores across all systems, judge: CLUE. Iter-B/G = Iter-QA-Base/Ginger, SS-B/G = SS-QA-Base/Ginger.}
\label{tab:CLUE-pair-2}
\end{table*}

\begin{table*}[]
\centering
\begin{minipage}{0.48\textwidth}
\centering

\begin{tabular}{ll|cc|cc}
\toprule
 & System & \multicolumn{2}{c|}{Info F1} & \multicolumn{2}{c}{Cite F1} \\
\cmidrule(lr){3-4} \cmidrule(lr){5-6}
 & & P & R & P & R \\
\midrule
\multirow{8}{*}{Q1}
    & CAG & 65.5 & \textbf{20.0} & 61.1 & 11.1 \\
\cmidrule{2-6}
    & Bullet & 56.2 & 15.6 & 56.2 & \textbf{13.3} \\
    & Ginger & 66.7 & \textbf{20.0} & \textbf{71.4} & 6.7 \\
    & RLM & 43.2 & 13.3 & 35.1 & \textbf{13.3} \\
    & Iter-B & 0.0 & 0.0 & 0.0 & 0.0 \\
    & Iter-G & 0.0 & 0.0 & 0.0 & 0.0 \\
    & SS-B & 1.7 & 2.2 & 1.7 & 4.4 \\
    & SS-G & \textbf{72.1} & 15.6 & 1.5 & 0.0 \\
\bottomrule
\end{tabular}
\subcaption{Central Texas Floods}
\label{tab:CLUE-central-texas-floods}
\end{minipage}
\hfill
\begin{minipage}{0.48\textwidth}
\centering

\begin{tabular}{ll|cc|cc}
\toprule
 & System & \multicolumn{2}{c|}{Info F1} & \multicolumn{2}{c}{Cite F1} \\
\cmidrule(lr){3-4} \cmidrule(lr){5-6}
 & & P & R & P & R \\
\midrule
\multirow{8}{*}{Q1}
    & CAG & 70.0 & 69.2 & 63.2 & 56.4 \\
\cmidrule{2-6}
    & Bullet & 63.6 & 61.5 & 59.1 & 59.0 \\
    & Ginger & 63.2 & 69.2 & 54.1 & 56.4 \\
    & RLM & 44.8 & 61.5 & 41.4 & 38.5 \\
    & Iter-B & 60.9 & \textbf{76.9} & 47.8 & 71.8 \\
    & Iter-G & 75.0 & \textbf{76.9} & 66.7 & 69.2 \\
    & SS-B & 64.6 & 74.4 & 64.6 & \textbf{76.9} \\
    & SS-G & \textbf{75.9} & 64.1 & \textbf{67.9} & 59.0 \\
\midrule
\multirow{8}{*}{Q2}
    & CAG & 72.4 & 48.7 & 69.0 & 43.6 \\
\cmidrule{2-6}
    & Bullet & 64.3 & 51.3 & 57.1 & 48.7 \\
    & Ginger & 67.7 & 48.7 & \textbf{71.0} & 43.6 \\
    & RLM & \textbf{75.0} & 53.8 & 62.5 & 41.0 \\
    & Iter-B & 51.6 & 53.8 & 39.1 & \textbf{56.4} \\
    & Iter-G & 67.4 & 48.7 & 47.6 & 38.5 \\
    & SS-B & 64.4 & 53.8 & 55.2 & 53.8 \\
    & SS-G & 73.0 & \textbf{56.4} & 45.9 & 43.6 \\
\bottomrule
\end{tabular}
\subcaption{Liberation Day Tariffs}
\label{tab:CLUE-liberation-day-tariffs}
\end{minipage}
\caption{Per-query scores across all systems, judge: CLUE. Iter-B/G = Iter-QA-Base/Ginger, SS-B/G = SS-QA-Base/Ginger.}
\label{tab:CLUE-pair-3}
\end{table*}

\begin{table*}[]
\centering
\begin{minipage}{0.48\textwidth}
\centering

\begin{tabular}{ll|cc|cc}
\toprule
 & System & \multicolumn{2}{c|}{Info F1} & \multicolumn{2}{c}{Cite F1} \\
\cmidrule(lr){3-4} \cmidrule(lr){5-6}
 & & P & R & P & R \\
\midrule
\multirow{8}{*}{Q1}
    & CAG & 65.4 & 13.2 & 73.1 & 2.9 \\
\cmidrule{2-6}
    & Bullet & 55.6 & 10.3 & 66.7 & 2.9 \\
    & Ginger & 73.9 & 13.2 & 82.6 & 2.9 \\
    & RLM & \textbf{88.6} & \textbf{29.4} & \textbf{94.3} & \textbf{17.6} \\
    & Iter-B & 0.0 & 0.0 & 0.0 & 0.0 \\
    & Iter-G & 0.0 & 0.0 & 0.0 & 0.0 \\
    & SS-B & 13.0 & 2.9 & 6.7 & 1.5 \\
    & SS-G & 35.7 & 11.8 & 3.6 & 0.0 \\
\midrule
\multirow{8}{*}{Q2}
    & CAG & \textbf{95.2} & 32.4 & \textbf{100.0} & 7.4 \\
\cmidrule{2-6}
    & Bullet & 89.5 & \textbf{38.2} & 84.2 & \textbf{23.5} \\
    & Ginger & 92.9 & 32.4 & \textbf{100.0} & 7.4 \\
    & RLM & 90.6 & 22.1 & 90.6 & 4.4 \\
    & Iter-B & 0.0 & 0.0 & 0.0 & 0.0 \\
    & Iter-G & 0.0 & 0.0 & 0.0 & 0.0 \\
    & SS-B & 8.5 & 2.9 & 6.9 & 1.5 \\
    & SS-G & 56.4 & 4.4 & 0.0 & 0.0 \\
\bottomrule
\end{tabular}
\subcaption{Nepal Youth Protests}
\label{tab:CLUE-nepal-youth-protests}
\end{minipage}
\hfill
\begin{minipage}{0.48\textwidth}
\centering

\begin{tabular}{ll|cc|cc}
\toprule
 & System & \multicolumn{2}{c|}{Info F1} & \multicolumn{2}{c}{Cite F1} \\
\cmidrule(lr){3-4} \cmidrule(lr){5-6}
 & & P & R & P & R \\
\midrule
\multirow{8}{*}{Q1}
    & CAG & \textbf{96.9} & 8.5 & 86.7 & 2.1 \\
\cmidrule{2-6}
    & Bullet & 88.9 & 10.1 & 88.9 & 1.6 \\
    & Ginger & \textbf{96.9} & 8.5 & \textbf{90.6} & 2.1 \\
    & RLM & 93.3 & 16.9 & 86.7 & 8.5 \\
    & Iter-B & 82.4 & \textbf{18.5} & 70.6 & 4.8 \\
    & Iter-G & 75.0 & 12.7 & 69.4 & 6.3 \\
    & SS-B & 56.0 & 18.0 & 54.0 & \textbf{11.6} \\
    & SS-G & 58.0 & 13.8 & 52.0 & 5.8 \\
\midrule
\multirow{8}{*}{Q2}
    & CAG & 97.7 & 7.9 & 95.2 & 2.1 \\
\cmidrule{2-6}
    & Bullet & 94.4 & 9.0 & \textbf{100.0} & 3.2 \\
    & Ginger & \textbf{98.0} & 7.9 & 93.8 & 2.1 \\
    & RLM & 90.9 & 12.2 & 81.8 & 5.3 \\
    & Iter-B & 80.3 & 13.8 & 78.7 & 10.6 \\
    & Iter-G & 67.4 & 13.8 & 52.2 & 5.3 \\
    & SS-B & 78.0 & \textbf{15.3} & 71.2 & \textbf{11.6} \\
    & SS-G & 85.1 & 12.2 & 68.1 & 7.4 \\
\bottomrule
\end{tabular}
\subcaption{Palisades Fire}
\label{tab:CLUE-palisades-fire}
\end{minipage}
\caption{Per-query scores across all systems, judge: CLUE. Iter-B/G = Iter-QA-Base/Ginger, SS-B/G = SS-QA-Base/Ginger.}
\label{tab:CLUE-pair-4}
\end{table*}

\begin{table*}[]
\centering
\begin{minipage}{0.48\textwidth}
\centering

\begin{tabular}{ll|cc|cc}
\toprule
 & System & \multicolumn{2}{c|}{Info F1} & \multicolumn{2}{c}{Cite F1} \\
\cmidrule(lr){3-4} \cmidrule(lr){5-6}
 & & P & R & P & R \\
\midrule
\multirow{8}{*}{Q1}
    & CAG & 62.9 & 37.9 & 57.1 & 12.9 \\
\cmidrule{2-6}
    & Bullet & 63.6 & 41.7 & 66.7 & \textbf{35.6} \\
    & Ginger & 61.1 & 37.9 & 58.3 & 12.9 \\
    & RLM & 70.0 & \textbf{46.2} & \textbf{70.0} & 28.0 \\
    & Iter-B & 74.6 & 25.8 & 3.4 & 0.0 \\
    & Iter-G & \textbf{80.8} & 24.2 & 7.7 & 0.0 \\
    & SS-B & 17.9 & 9.8 & 10.3 & 3.0 \\
    & SS-G & 64.8 & 26.5 & 57.4 & 0.0 \\
\midrule
\multirow{8}{*}{Q2}
    & CAG & 73.0 & 37.9 & 70.3 & 18.9 \\
\cmidrule{2-6}
    & Bullet & 76.0 & \textbf{43.2} & 84.0 & \textbf{35.6} \\
    & Ginger & 76.5 & 37.9 & 70.6 & 18.9 \\
    & RLM & \textbf{78.9} & 42.4 & \textbf{92.1} & 18.2 \\
    & Iter-B & 0.0 & 0.0 & 0.0 & 0.0 \\
    & Iter-G & 0.0 & 0.0 & 0.0 & 0.0 \\
    & SS-B & 2.4 & 5.3 & 9.5 & 3.8 \\
    & SS-G & 77.4 & 15.9 & 1.6 & 0.0 \\
\bottomrule
\end{tabular}
\subcaption{Shi Yongxin Scandal}
\label{tab:CLUE-shi-yongxin-scandal}
\end{minipage}
\hfill
\begin{minipage}{0.48\textwidth}
\centering

\begin{tabular}{ll|cc|cc}
\toprule
 & System & \multicolumn{2}{c|}{Info F1} & \multicolumn{2}{c}{Cite F1} \\
\cmidrule(lr){3-4} \cmidrule(lr){5-6}
 & & P & R & P & R \\
\midrule
\multirow{8}{*}{Q1}
    & CAG & 79.3 & \textbf{28.8} & 79.3 & \textbf{23.5} \\
\cmidrule{2-6}
    & Bullet & 76.9 & \textbf{28.8} & 80.8 & 14.2 \\
    & Ginger & \textbf{85.7} & 25.6 & \textbf{81.6} & 18.5 \\
    & RLM & \textbf{85.7} & 22.8 & 68.6 & 17.4 \\
    & Iter-B & 37.7 & 10.0 & 15.1 & 2.1 \\
    & Iter-G & 29.5 & 9.3 & 13.6 & 3.2 \\
    & SS-B & 32.3 & 8.9 & 29.0 & 3.9 \\
    & SS-G & 45.7 & 7.1 & 25.0 & 3.6 \\
\midrule
\multirow{8}{*}{Q2}
    & CAG & 86.2 & 15.3 & 75.9 & 5.0 \\
\cmidrule{2-6}
    & Bullet & 84.6 & 15.7 & \textbf{84.6} & 8.2 \\
    & Ginger & \textbf{90.3} & 15.3 & 83.3 & 5.0 \\
    & RLM & 68.6 & \textbf{22.4} & 62.9 & \textbf{10.3} \\
    & Iter-B & 0.0 & 0.0 & 0.0 & 0.0 \\
    & Iter-G & 0.0 & 0.0 & 0.0 & 0.0 \\
    & SS-B & 3.2 & 6.0 & 3.2 & 1.1 \\
    & SS-G & 11.9 & 7.5 & 10.6 & 1.1 \\
\bottomrule
\end{tabular}
\subcaption{Tropical Storm Wipha}
\label{tab:CLUE-tropical-storm-wipha}
\end{minipage}
\caption{Per-query scores across all systems, judge: CLUE. Iter-B/G = Iter-QA-Base/Ginger, SS-B/G = SS-QA-Base/Ginger.}
\label{tab:CLUE-pair-5}
\end{table*}
Tables~\ref{tab:CLUE-pair-1}, \ref{tab:CLUE-pair-2}, \ref{tab:CLUE-pair-3}, \ref{tab:CLUE-pair-4}, and~\ref{tab:CLUE-pair-5} report per-query scores across all systems and topics. Claim-based systems (CAG, Bullet, Ginger) consistently achieve high information precision but lower recall. QA systems suffer catastrophic zero-score failures on several topics, including the Alaskan Typhoon (\autoref{tab:CLUE-2025-alaskan-typhoon}), Central Texas Floods (\autoref{tab:CLUE-central-texas-floods}), Nepal Youth Protests (\autoref{tab:CLUE-nepal-youth-protests}), and Shi Yongxin Scandal (\autoref{tab:CLUE-shi-yongxin-scandal}), where the QA pipeline fails to retrieve relevant videos and the generator backs off rather than hallucinate. The RLM performs most distinctively on the Canadian Federal Election (\autoref{tab:CLUE-2025-canadian-federal-election}), where cross-video conflict resolution yields the highest citation precision and recall on Q2, and on Nepal Youth Protests, where it substantially outperforms all other systems. The Palisades Fire (\autoref{tab:CLUE-palisades-fire}) and Tropical Storm Wipha (\autoref{tab:CLUE-tropical-storm-wipha}) exhibit uniformly high precision but very low recall across all systems, suggesting broad reference sets that no system fully covers. 

In \autoref{tab:leaderboard-retrieval-snapshot} and \autoref{tab:leaderboard-oracle-snapshot} we report the overall rankings of each system from the MAGMaR shared task leaderboards for retrieval and generation, respectively. Our retrieval systems hold the 2nd-6th place positions. Our generation systems place 1st and 3rd-6th.

\begin{table*}[t]
\centering
\setlength{\tabcolsep}{3pt}
\begin{tabular}{l|l|ccccccc}
\toprule
\textbf{Rank} & \textbf{System} & \textbf{Avg.} & \textbf{nDCG@10} & \textbf{nDCG@20} & \textbf{nDCG@100} & \textbf{R@10} & \textbf{R@20} & \textbf{R@100} \\
\midrule
2 & RRF K=10 + RV & 0.759 & 0.759 & 0.771 & 0.811 & 0.652 & 0.735 & 0.832 \\
3 & Weighted RRF + RV & 0.757 & 0.757 & 0.768 & 0.810 & 0.650 & 0.725 & 0.832 \\
4 & RRF K=60 + RV & 0.751 & 0.754 & 0.765 & 0.807 & 0.641 & 0.716 & 0.823 \\
5 & Sum Sim + RV & 0.745 & 0.747 & 0.758 & 0.800 & 0.636 & 0.711 & 0.818 \\
6 & RRF K=100 + RV & 0.744 & 0.746 & 0.757 & 0.799 & 0.636 & 0.711 & 0.818 \\
\bottomrule
\end{tabular}
\caption{MAGMaR retrieval final leaderboard positions for MARS submissions. Rank is the public leaderboard rank under the default average over the six displayed retrieval metrics. }
\label{tab:leaderboard-retrieval-snapshot}
\end{table*}

\begin{table*}[t]
\centering
\setlength{\tabcolsep}{3pt}
\begin{tabular}{l|l|ccc|cc|cc}
\toprule
\textbf{Rank} & \textbf{System} & \textbf{Human} & \textbf{Best Votes} & \textbf{Best \%} & \textbf{Info P} & \textbf{Info R} & \textbf{Cite P} & \textbf{Cite R} \\
\midrule
1  & \iterqabase{}   & 3.833 & 8  & 14.0\% & 0.347 & 0.313 & 0.268 & 0.258 \\
3  & \iterqaginger{} & 3.694 & 5  & 8.8\%  & 0.345 & 0.290 & 0.257 & 0.226 \\
4  & \ssqaginger{}   & 3.421 & 10 & 17.5\% & 0.544 & 0.324 & 0.326 & 0.238 \\
5  & \marquis-\rlm{}           & 3.298 & 3  & 5.3\%  & 0.708 & 0.385 & 0.592 & 0.272 \\
6  & \ginger{}         & 3.123 & 6  & 10.5\% & 0.776 & 0.404 & 0.643 & 0.226 \\
8  & \ssqabase{}    & 3.070 & 6  & 10.5\% & 0.331 & 0.306 & 0.277 & 0.281 \\
10 & \bulletmethod{}        & 2.667 & 0  & 0.0\%  & 0.711 & 0.394 & 0.604 & 0.237 \\
\bottomrule
\end{tabular}
\caption{MAGMaR oracle article-generation leaderboard snapshot for \marquis{} submissions. Rank is the public leaderboard rank under the default Human Score ordering.}
\label{tab:leaderboard-oracle-snapshot}
\end{table*}

\section{Appendix: Retrieval Implementation and Full Ablation}
\label{app:retrieval}
This appendix consolidates the retrieval implementation details, query
expansion artifacts, retrieval figure, ablations, and hyperparameters. Prompt
templates for query decomposition are listed in \autoref{app:prompts}.



\subsection{Query and Corpus Encoding}
\label{app:retrieval-encoding}

Each MAGMaR query is represented by concatenating the persona title,
background, and query text. The original queries and generated sub-queries are
encoded with OmniEmbed using the same query prefix, appended end-of-text token,
end-of-sequence pooling, and L2 normalization. The video corpus contains
109,814 videos, consisting of 109,724 MultiVENT 2.0 videos and 90 MAGMaR2026
test videos. Search uses cosine similarity over normalized embeddings and
returns the top 100 videos per query or sub-query.

\subsection{Sub-query Expansion Statistics}
\label{app:retrieval-expansion-stats}

The final flattened sub-query file contains 430 sub-queries across 19 original
queries, for an average of 22.63 sub-queries per query. The minimum is 1 and
the maximum is 25. The minimum is caused by one malformed decomposition output
that fell back to a single query-like search probe in the flattened retrieval
file. Excluding this fallback case, the 18 successfully decomposed queries
produce 429 sub-queries, with a minimum of 22, an average of 23.83, and a
maximum of 25 sub-queries per query.

\begin{table*}[t]
\centering
\begin{tabular}{lccccc}
\toprule
Set & Queries & Total & Min & Avg. & Max \\
\midrule
Final flattened retrieval file & 19 & 430 & 1 & 22.63 & 25 \\
Successful decompositions only & 18 & 429 & 22 & 23.83 & 25 \\
\bottomrule
\end{tabular}
\caption{Sub-query expansion statistics.}
\label{tab:retrieval-expansion-stats}
\end{table*}

\subsection{Qualitative Expansion Examples}
\label{app:retrieval-expansion-examples}
In \autoref{tab:retrieval-expansion-examples} we show some example expansions from our method. 

\begin{table*}[t]
\centering
\begin{tabular}{p{0.45\textwidth}p{0.45\textwidth}}
\toprule
\textbf{2025 Canadian federal election} & \textbf{2025 Alaska typhoon} \\
\midrule
2025 Canadian federal election final seat count by party &
2025 Alaska typhoon housing damage assessment report \\
Elections Canada 2025 election official results dataset &
residential structural failures Alaska 2025 typhoon \\
2025 Canadian federal election popular vote share by party &
housing construction materials damaged 2025 Alaska storm \\
2025 Canadian federal election seat changes by party &
geographic distribution of typhoon damage Alaska 2025 \\
2025 Canadian federal election candidate vote totals &
roof failure mechanisms 2025 Alaska coastal storm \\
voter turnout rate Canada 2025 vs 2021 &
foundation erosion damage coastal Alaska 2025 \\
\bottomrule
\end{tabular}
\caption{Representative sub-queries produced by the query decomposition stage.}
\label{tab:retrieval-expansion-examples}
\end{table*}

\subsection{Fusion and Reranking}
\label{app:retrieval-fusion}

We evaluate both score-based and rank-based fusion over the sub-query ranked
lists. The score-based methods are sum similarity, max similarity, and mean
similarity. The rank-based methods are reciprocal rank fusion with
$K\in\{10,60,100\}$ and weighted reciprocal rank fusion, where reciprocal-rank
contributions are weighted by cosine similarity. We also evaluate reranked
variants in which the top 100 first-stage candidates are reordered with
\rankvideo.


\subsection{Dropping Sub-queries}
\label{app:retrieval-drop-ablation}

To test whether query decomposition depends on a small number of strong
sub-queries or on broad facet coverage, we randomly retain only $k$ sub-queries
per original query before fusion, for $k~\in~\{1,5,10\}$. We repeat each random
condition over five seeds and report mean and standard deviation. The full
system uses all generated sub-queries.

\begin{table*}[t]
\centering
\begin{tabular}{lcccc}
\toprule
Sub-queries kept & nDCG@10 & nDCG@100 & R@10 & R@100 \\
\midrule
1 random & 0.613 $\pm$ 0.029 & 0.679 $\pm$ 0.025 & 0.512 $\pm$ 0.027 & 0.707 $\pm$ 0.024 \\
5 random & 0.684 $\pm$ 0.034 & 0.749 $\pm$ 0.029 & 0.601 $\pm$ 0.041 & 0.794 $\pm$ 0.031 \\
10 random & 0.696 $\pm$ 0.010 & 0.763 $\pm$ 0.013 & 0.615 $\pm$ 0.018 & 0.812 $\pm$ 0.015 \\
All & 0.711 & 0.773 & 0.640 & 0.831 \\
\bottomrule
\end{tabular}
\caption{Effect of randomly retaining fewer sub-queries before max-similarity
fusion. Random conditions are averaged over five seeds.}
\label{tab:retrieval-drop-ablation}
\end{table*}

Performance improves monotonically as more sub-queries are retained. A single
random sub-query already outperforms the no-expansion baseline, showing that
the decomposition often produces useful search probes. However, retaining 5
or 10 sub-queries substantially improves both ranking quality and recall, and
using all sub-queries gives the best overall performance. This suggests that
the gains from decomposition come not only from finding one strong reformulation
but also from covering multiple facets of the original query.

\section{Appendix: Information Extraction Implementation}

\label{app:information-extraction}
\begin{figure*}[t]
    \centering
    \includegraphics[width=\textwidth]{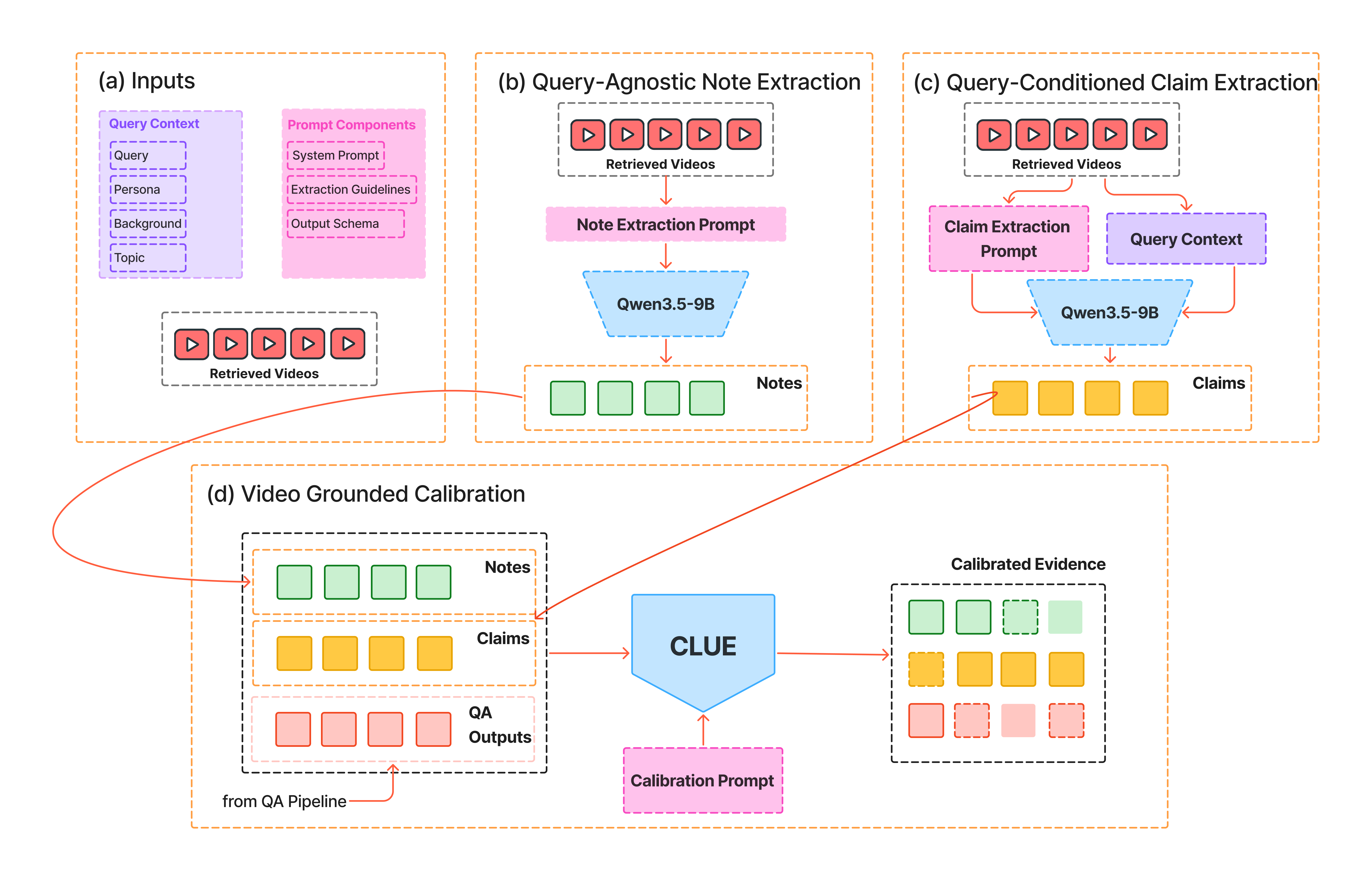}
    \caption{Information extraction and calibration workflow. Retrieved videos and prompt components are used to produce query-agnostic notes and query-conditioned claims, while QA outputs enter from the question-answering pipeline (See \autoref{app:qa} and \autoref{fig:qa}). These extracted evidence records are merged and scored against source video by the calibration backend, producing calibrated extracted evidence for article generation.}
    \label{fig:info-extraction}
\end{figure*}

This appendix provides implementation details for the information extraction stage, including general note extraction, query-conditioned claim extraction, artifact schemas, representative outputs, and query--topic alignment. Prompt templates for note extraction, claim extraction, and calibration are listed in \autoref{app:prompts}.
\autoref{fig:info-extraction} illustrates the information extraction and calibration workflow.

\subsection{General Note Extraction}

General note extraction is run independently for each video. The extractor receives the video together with topic and video metadata and produces atomic observations describing directly observable visual content, OCR, and spoken or audio evidence. The output is a JSON object containing a list of notes. Each note includes note text, modality, and an optional timestamp.

\subsection{Query-Conditioned Claim Extraction}

Query-conditioned claim extraction is run for query--video pairs after aligning each evaluation query to a topic-specific video subset. The extractor receives the query identifier, topic, persona title, background, query text, and video identifier, and outputs claims relevant to the information need. Each claim is tied to a specific query and video and may include confidence, evidence description, source type, and timestamp metadata.

\subsection{Query--Topic Alignment}

The official query set contains 19 evaluation queries. Each query is aligned to one of 10 topic buckets through deterministic title-to-topic normalization. Query-conditioned claim extraction is then applied over the videos mapped to the corresponding topic. General note extraction uses the topic identity only as metadata and does not condition on the evaluation query.

\subsection{Extracted Evidence Schemas}

A general note contains a note identifier, video identifier, topic label, note text, modality tag, and optional timestamp. A query-conditioned claim contains a claim identifier, query identifier, video identifier, topic label, claim text, and optional support-oriented metadata such as confidence, evidence description, source type, and timestamp.

\subsection{Representative Outputs}

To make the extraction flow concrete, we show one representative output from each major stage. These examples are lightly trimmed for presentation but preserve the actual field structure used by the pipeline.

\paragraph{Example general note.}
\begin{examplebox}
\pline{\{}
\pline{\ \ "note\_id": "gn1a-hol6y3QwX2Y-000",}
\pline{\ \ "video\_id": "hol6y3QwX2Y",}
\pline{\ \ "topic": "2025\_Canadian\_Federal\_Election",}
\pline{\ \ "text": "A woman with short blonde hair and a beige jacket is speaking.",}
\pline{\ \ "modality": "visual",}
\pline{\ \ "timestamp": [0.0, 6.0]}
\pline{\}}
\end{examplebox}

\paragraph{Example query-conditioned claim.}
\begin{examplebox}
\pline{\{}
\pline{\ \ "claim\_id": "qc-10-1978302738418032640-000",}
\pline{\ \ "query\_id": "10",}
\pline{\ \ "video\_id": "1978302738418032640",}
\pline{\ \ "topic": "2025\_Alaska\_Typhoon",}
\pline{\ \ "claim": "More than 50 people have been rescued in Western Alaska.",}
\pline{\ \ "confidence": 0.95,}
\pline{\ \ "evidence": "Text overlay in the video states 'More than 50 people have been rescued in Western Alaska.'",}
\pline{\ \ "source": "video\_text",}
\pline{\ \ "timestamp": [0.0, 3.0]}
\pline{\}}
\end{examplebox}

\section{Appendix: Calibration Implementation}

\label{app:calibration}
This appendix provides implementation details for video-grounded calibration. Calibration is run after information extraction and assigns a support probability to each extracted artifact without modifying the original artifact content. The calibration prompt itself is provided in Appendix~\ref{app:prompts}.

\subsection{Calibration Inputs and Outputs}

For each extracted artifact, the calibration stage receives the source video and the artifact text. The output is a scalar support probability in $[0,1]$ estimating whether the artifact is supported by the source video. The calibrated artifact preserves the original note or claim and attaches a calibration payload containing the support score and backend provenance.

\subsection{Backends}

We evaluate two calibration backends. The prompted backend uses Qwen3.5 with a constrained probability-estimation prompt. The comparison backend is CLUE built on the Qwen2.5-Omni family. Both backends consume the same video--artifact pairs and emit the same conceptual output type.

\subsection{Attachment Logic}

Calibration predictions are attached to artifacts using stable artifact identifiers when available. When identifiers are unavailable or inconsistent, the attachment stage falls back to matching by video identifier and artifact text. This preserves compatibility across extraction and calibration jobs while keeping the original artifact representation unchanged.

\paragraph{Example calibrated artifact.}
\begin{examplebox}
\pline{\{}
\pline{\ \ "claim\_id": "qc-10-1978302738418032640-000",}
\pline{\ \ "claim": "More than 50 people have been rescued in Western Alaska.",}
\pline{\ \ "calibration": \{}
\pline{\ \ \ \ "unli": \{}
\pline{\ \ \ \ \ \ "prob": 0.95,}
\pline{\ \ \ \ \ \ "raw": \{}
\pline{\ \ \ \ \ \ \ \ "raw\_output": "\textless answer\textgreater 0.95 \textless /answer\textgreater"}
\pline{\ \ \ \ \ \ \}}
\pline{\ \ \ \ \}}
\pline{\ \ \}}
\pline{\}}
\end{examplebox}

\subsection{Claim Filtering}

In addition to attaching support probabilities, the calibration stage can optionally filter extracted claims using the predicted support score. Claims with support probabilities below a predefined threshold are excluded from downstream outputs. This filtering mechanism is intended to reduce unsupported or weakly grounded claims while preserving high-confidence artifacts.

The filtering threshold is configurable and applied uniformly across calibration backends. Importantly, filtering is performed only after extraction and does not modify the original extracted content or calibration predictions.

For the calibration models(CLUE and Qwen3.5), the hyperparameters are summarized in \autoref{tab:unified-hparams}.

\section{Appendix: Question Answering Implementation}

\label{app:qa}
\begin{figure*}
    \centering
    \includegraphics[width=\linewidth]{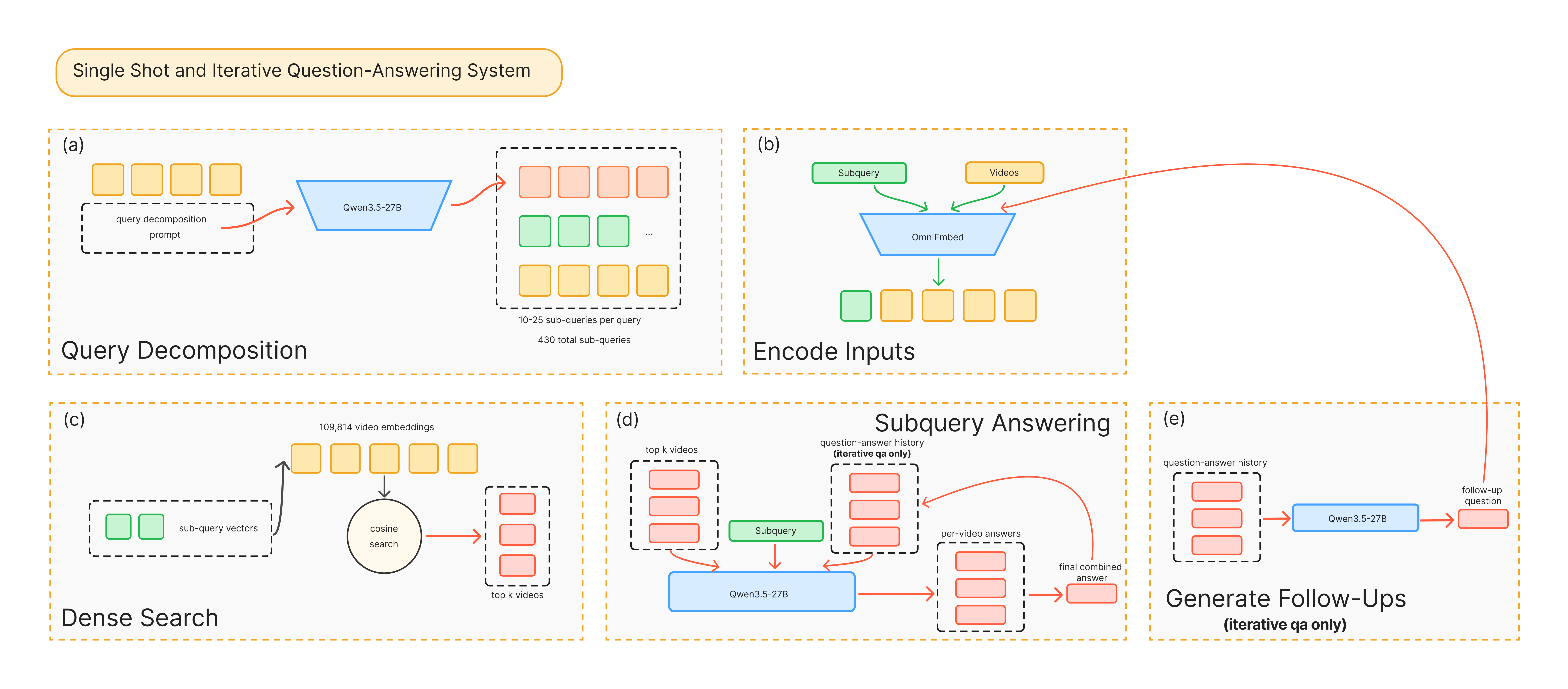}
    \caption{Overview of QA-based evidence extraction method. The single-shot variant decomposes the query into fixed subqueries, retrieves videos, answers each subquery, and aggregates the answers. The iterative variant generates follow-up questions from the question-answer history until a stopping condition is met.}
    \label{fig:qa}
\end{figure*}

Figure \ref{fig:qa} provides an overview of the single shot and iterative question answering systems.

\textbf{Single-Shot Question Answering. } The single-shot pipeline uses Qwen3.5-27B for answer generation, Qwen2.5-Omni-7B with OmniEmbed-v0.1 for multimodal embeddings, and Whisper (medium.en) for transcription. 

Videos are preprocessed by downsampling to 30 FPS with a max frames of 32. Transcripts are generated from the audio stream and paired with each video. Video and query embeddings are computed in a shared space using the OmniEmbed model, and retrieval is performed using cosine similarity with a threshold of 0.1 and top-k selection with k=4.

For each sub-query, the system retrieves relevant videos and generates per-video answers using Qwen3.5-27B, conditioned jointly on the video frames, transcript, and query. The model is prompted to produce concise factual answers grounded only in the provided inputs. All per-video answers are collected, and responses such as ``I don't know'' are filtered out. The remaining answers are then merged using a second Qwen3.5-27B pass that combines the extracted answers into a single response without introducing external knowledge.

Across the evaluated queries, the single-shot pipeline generated on average 23.84 expanded questions per main query, with a minimum of 20 and a maximum of 25 expanded questions. In the single-shot setting, 246 of 453 expanded questions were unanswered. Qualitative examples from the single-shot output for the first main query are shown below.

\begin{table}[t]
\centering
\small
\begin{tabular}{p{0.45\linewidth} p{0.45\linewidth}}
\hline
\textbf{Expanded question} & \textbf{Answer} \\
\hline
What were the final seat counts for each political party in the 2025 Canadian federal election? &
Based on the provided video transcript with 96\% of votes counted, the projected seat counts were Liberals: 166, Conservatives: 146, Bloc Qu\'eb\'ecois: 23, and NDP: 7. The final official counts after 100\% of votes were counted were unknown based on the provided text. \\
\hline
What is the official dataset published by Elections Canada containing detailed results for the 2025 federal election? &
I don't know. The specific name of the official dataset published by Elections Canada containing detailed results for the 2025 federal election is unknown based on the provided information. \\
\hline
What were the popular vote percentages for the Liberal Party and Conservative Party in the 2025 Canadian federal election? &
The Liberal Party received 43.1\% of the popular vote and the Conservative Party received 41.7\%. \\
\hline
\end{tabular}
\caption{Examples of single-shot expanded questions and answers for the first main query.}
\end{table}

\textbf{Iterative Question Generation. } The iterative pipeline uses the same underlying structure as the single-shot approach, so each step performs retrieval and per-video answer generation in the same way as the single-shot pipeline.

The key difference is that instead of processing each sub-query once, the system maintains a running history of question--answer pairs and iteratively refines the query. After each retrieval and answer aggregation step, the aggregated answer is appended to the history, and a new follow-up question is generated using Qwen3.5-27B with sampling enabled, prompting the model to produce exactly one question that extracts additional or more specific information conditioned on the full history. This loop continues for up to 5 steps per sub-query but terminates early if no videos are retrieved, the model outputs ``NONE'' as the next question, or a repeated question is detected.

Across the evaluated queries, the iterative pipeline generated a minimum of 22 expanded questions, a maximum of 73 expanded questions, and an average of 41.05 expanded questions per main query. In the iterative setting, 293 of 613 expanded questions were unanswered.

Qualitative examples from the iterative output for the first main query are shown below. These examples illustrate how the iterative method can generate follow-up questions that become more specific than the original expanded questions.

\begin{table}[h]
\centering
\small
\begin{tabular}{p{0.45\linewidth} p{0.45\linewidth}}
\hline
\textbf{Expanded question} & \textbf{Answer} \\
\hline
What were the final seat counts for each political party in the 2025 Canadian federal election? &
Based on the provided transcript with 96\% of votes counted, the projected seat counts were Liberals: 166, Conservatives: 146, Bloc Qu\'eb\'ecois: 23, and NDP: 7. The final official seat counts were unknown because the provided data represented projections before 100\% of votes were counted. \\
\hline
Which specific electoral districts accounted for the largest swing in votes that resulted in the projected reduction of the NDP's seat count to seven? &
Fortress Vancouver, Fortress Montreal, and the GTA accounted for the largest swing in votes that resulted in the projected reduction of the NDP's seat count to seven. \\
\hline
\end{tabular}
\caption{Examples of iterative expanded questions and answers for the first main query.}
\end{table}

\section{Appendix: Article Generation Implementation}

\label{app:article-generation}

This appendix provides implementation details for the article generation systems. Prompt templates for all article generation variants are provided in Appendix~\ref{app:prompts}.

\subsection{Evidence Inputs}

The article generation systems operate over flat lists of evidence artifacts. These artifacts may be query-conditioned claims, query-agnostic notes, or QA pairs. Claims and notes include video identifiers and timestamps when available. QA pairs include the source videos used to produce the answer.

\subsection{\bulletmethod{} Generation}

The bullet-point generator renders selected evidence items directly as a numbered list of findings with inline citations. This variant does not invoke a generation model and is intended as a conservative evidence-presentation baseline.

\subsection{Single-Prompt Article Generation}

The single-prompt article generator concatenates the evidence items for a query into a single prompt and generates a coherent article with inline citations. To reduce context length and memory failures, evidence sets larger than 25 items are truncated to the top 25 by confidence score.

\subsection{\ginger{} Article Generation}

The \ginger{} generator decomposes article generation into facet clustering, cluster ranking, per-cluster summarization, and fluency enhancement. Since the information extraction stage already produces atomic evidence units, the pipeline begins from extracted notes, claims, or QA pairs rather than running a separate nugget-detection stage.

\section{Appendix: RLM Controller Implementation}

\label{app:rlm}

\lstdefinelanguage{myjson}{
    basicstyle=\footnotesize\ttfamily,
    showstringspaces=false,
    breaklines=true,
    string=[s]{"}{"},
    stringstyle=\color{jsonstr},
    keywords={true,false,null},
    keywordstyle=\color{jsonkw}\bfseries,
}

\newtcblisting{jsonbox}{%
  breakable,
  enhanced,
  colback=gray!4,
  colframe=black!35,
  boxrule=0.5pt,
  arc=1mm,
  left=6pt,
  right=6pt,
  top=6pt,
  bottom=6pt,
  listing only,
  listing options={language=myjson},
}
\lstdefinestyle{agentlogstyle}{
    basicstyle=\fontsize{7.5}{8}\selectfont\ttfamily,
    breaklines=true,
    breakatwhitespace=false,
    breakautoindent=false,
    breakindent=0pt,
    keepspaces=false,
    showstringspaces=false,
    columns=fixed,
    basewidth=0.52em,
    upquote=true,
    aboveskip=2pt,
    belowskip=2pt,
}

\newtcblisting{agentlog}{%
  breakable,
  enhanced,
  colback=black!2,
  colframe=black!30,
  boxrule=0.3pt,
  arc=0pt,
  left=1pt,
  right=1pt,
  top=2pt,
  bottom=2pt,
  listing only,
  listing options={style=agentlogstyle},
}
\label{sec:app-rlm}

\begin{figure*}[t]
    \centering
    \includegraphics[width=\textwidth]{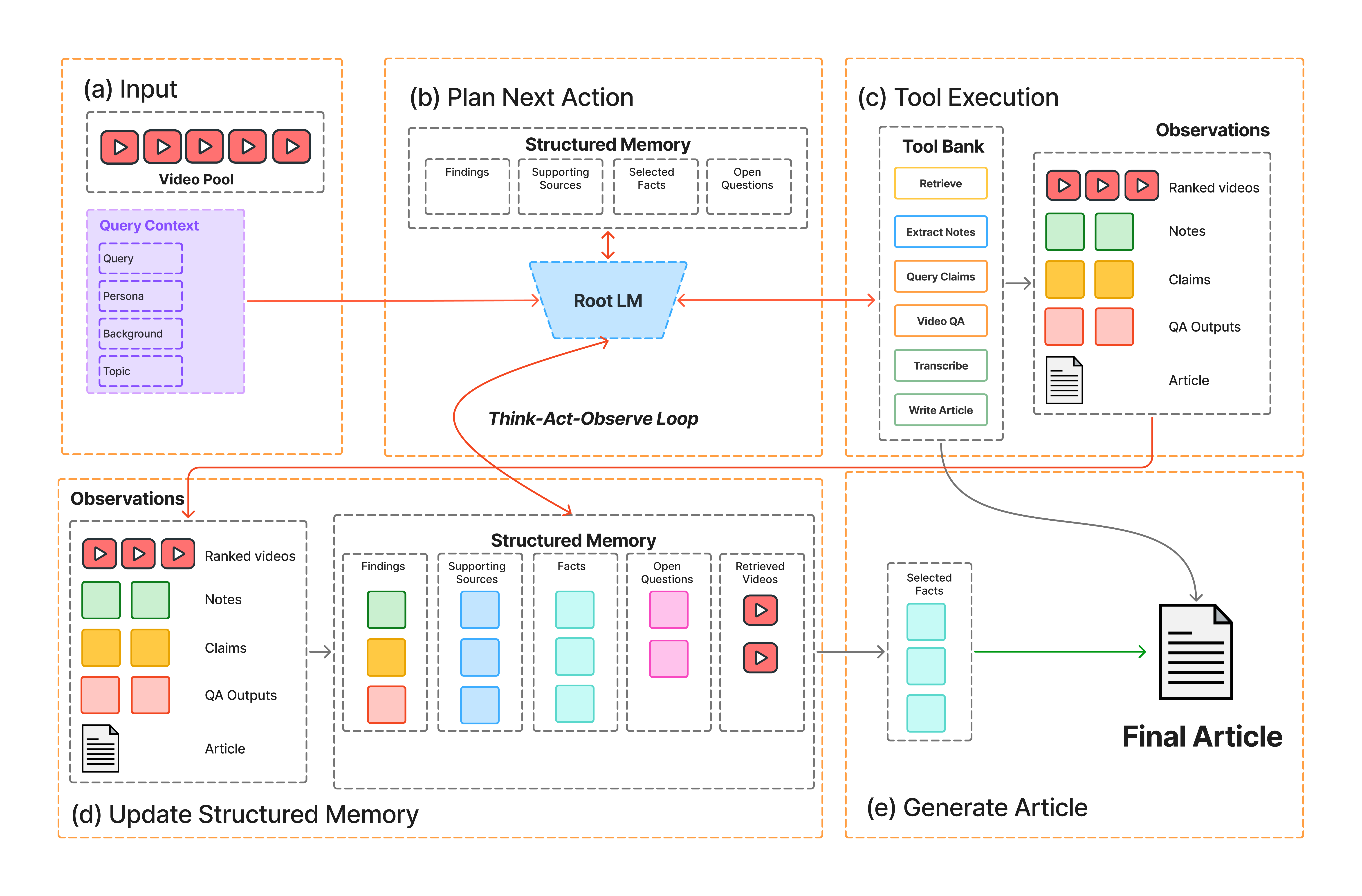}
    \caption{\marquis-\rlm{} controller. The \rootlm{} reads structured memory, plans the next action, executes one tool call in a persistent REPL environment, observes the result, and updates memory before continuing. Once sufficient evidence has been gathered and judged, selected facts are passed to the article-generation tool to produce the final cited article.}
    \label{fig:rlm-controller}
\end{figure*}

This appendix documents the tool API, memory schema, and prompts used by the RLM controller (see \autoref{sec:rlm} in the main text). The goal is to make the RLM-side of our submission directly reproducible. \autoref{fig:rlm-controller} summarizes the resulting Think--Act--Observe control loop.

\subsection{Tool API and Backing Modules}

\autoref{tab:rlm_tools} lists all callable tool functions registered in the REPL namespace.

\begin{table*}
\centering
\small
\setlength{\tabcolsep}{3pt}
\begin{tabular}{p{0.20\textwidth} p{0.34\textwidth} p{0.40\textwidth}}
\toprule
\textbf{Tool} & \textbf{Signature} & \textbf{Backing module} \\
\midrule
\multicolumn{3}{l}{\textit{Perception tools (multimodal extraction over raw video)}} \\
Caption        & \textit{video\_caption(vid)}        & local Qwen3.5-9B model \\
GeneralNotes   & \textit{general\_notes(vid)}        & \autoref{sec:info-ext-general-notes} general note extraction  \\
QueryClaims    & \textit{query\_claims(vid)}         & \autoref{sec:info-ext-query-claims} query-conditioned claim extraction \\
\midrule
\multicolumn{3}{l}{\textit{Targeted-query tools}} \\
VideoQA        & \textit{video\_qa(vid, question)}   & \autoref{sec:info-ext-qa} multimodal QA pipeline \\
Transcribe     & \textit{transcribe(vid)}            & local Whisper Model \\
RetrieveChunks & \textit{retrieve\_chunks(vid)}      & \autoref{sec:retrieval} OmniEmbed retriever, lowered to 20\,s chunk level \\
\midrule
\multicolumn{3}{l}{\textit{Generation tool}} \\
WriteReport    & \textit{write\_report(facts)}       & \autoref{sec:report_generation} GINGER-based pipeline \\
\midrule
\multicolumn{3}{l}{\textit{Memory operators}} \\
memory\_summary     & \textit{memory\_summary()}             & compact memory snapshot per iteration \\
print\_memory       & \textit{print\_memory(slot=None)}      & full JSON dump of one or all slots \\
add\_keyword        & \textit{add\_keyword(vid, kw)}         & tag a video with a keyword \\
search\_by\_keyword & \textit{search\_by\_keyword(kw)}       & find items in memory bank \\
remove\_fact        & \textit{remove\_fact(vid, idx)}        & delete a single fact \\
clear\_facts        & \textit{clear\_facts(vid=None)}        & clear all facts for one or all videos \\
\midrule
\multicolumn{3}{l}{\textit{Memory operators}} \\
Think          & \textit{llm\_think()}   & fact\_table $\to$ findings (infered by LLM) \\
Judge          & \textit{llm\_judge()}   & fact\_table $\to$ selected\_facts (for report generation) \\
\bottomrule
\end{tabular}
\caption{Tools and memory operators registered in the RLM REPL namespace.}
\label{tab:rlm_tools}
\end{table*}

\subsection{Memory Bank JSON Structure}

\begin{jsonbox}
{
  "findings":      [" high-level insights "],
  "keywords":      {"<video_id>": ["keyword1", "keyword2"]},
  "fact_table":    {"<video_id>": [
    {"fact": "...", "timestamp": "10s-15s", "source_tool": "query_claims", "confidence": 0.8}
  ]},
  "selected_facts": ["facts chosen by llm_judge for the final report"],
  "videos":        {"<video_id>": {"status": "processed", "tools_used": [...], "path": "...", "caption": "..."}}
}
\end{jsonbox}


\subsection{Examples of \rootlm{} Behavior}
\label{sec:rlm_observation}

\paragraph{Example 1 Final recheck.}

\begin{agentlog}

[THINK] (6.5s):
  The report has been generated but gaps remain. I should verify one final critical issue - whether the seat counts (Lib 120, CON 87) shown in zaFtBz84Kyk were final or still live projections, since this affects report accuracy. Let me verify this with video_qa one more time.

[ACT]:
  ```repl
  video_qa("zaFtBz84Kyk",
           "Was the seat count shown (Lib 120, CON 87) the final result or a live projection? Clarify the exact time shown and when results were finalized.")
  ```

In [42]:
  1  video_qa("zaFtBz84Kyk", "Was the seat count shown (Lib 120,
              CON 87) the final result or a live projection? ...")

Out [42]:
  '[video_qa] zaFtBz84Kyk - Q: Was the seat count shown (Lib 120, CON 87) the final result or a live projection? Clarify the exact time shown and when results were finalized. A: The seat count shown (Lib 120, CON 87) was a live projection, not the final result, as the final declaration occurred later with the banner "LIB WIN" and the count updated to 124 for Liberals. confidence=1.0, timestamp=220-230s'

[OBSERVE]:
  TOOL CALLS (1):
    OK  video_qa({'video_path': 'zaFtBz84Kyk', ...})  ->  82.7s

[JUDGE] iter 41 (model: qwen/qwen3.5-9b):
  Reasoning 5 | Action 5 | Granularity 5 | Progress 5
  Redundancy 5 | Think 5 | Code 5 | Output 5 | Tool 5
  WEIGHTED TOTAL: 62.5/62.5 (100.0%)
\end{agentlog}

\noindent\textbf{Observation.} \rootlm{} re-engages \texttt{video\_qa} to disambiguate a data point to prioritize correctness over completion.

\paragraph{Example 2 Vague Information Clarification.}

\begin{agentlog}

[THINK] (3.4s):
  I've watched all 17 videos but only queried 4 for facts. I'm missing critical casualty and aid distribution data. Video 1978302738418032640 caption mentions "at least one is dead and several remain missing" - I need to extract these numbers via video_qa.

[ACT]:
  ```repl
  video_qa("1978302738418032640","How many fatalities and missing persons are there?What details about rescue operations?")
  ```

In [27]:
  1  video_qa("1978302738418032640", "How many fatalities and missing persons are there? ...")

Out [27]:
  '[video_qa] 1978302738418032640 - Q: How many fatalities and missing persons are there? What details about rescue operations?
   A: At least one fatality and several missing; over 50 rescued via helicopter.
confidence=1.0, timestamp=8-15s'

[OBSERVE]:
  TOOL CALLS (1):
    OK  video_qa({'video_path': '1978302738418032640', ...})  ->  22.2s

[JUDGE] iter 26 (model: qwen/qwen3.5-9b):
  Reasoning 5 | Action 5 | Granularity 5 | Progress 5
  Redundancy 5 | Think 5 | Code 5 | Output 5 | Tool 5
  WEIGHTED TOTAL: 62.5/62.5 (100.0\%)
\end{agentlog}

\noindent\textbf{Observation.} \rootlm{} identified the information from the caption was vague and decide to call the QA tool to get a more precise answer.

\paragraph{Example 3 Cross-modal Conflict Resolution.}

\begin{agentlog}

[THINK] (4.1s):
  I need to resolve the major conflict - seat counts vary between r2P6c5tzvM8 (LIB 152, CON 120) and zaFtBz84Kyk (LIB 119-120, CON 84-87). r2P6c5tzvM8 appears more authoritative as CTV News official report. I should use transcribe to get the final declared results from both CTV videos to reconcile this conflict.

[ACT]:
  ```repl
  transcribe("r2P6c5tzvM8")
  ```

In [17]:
  1  transcribe("r2P6c5tzvM8")

Out [17]:
  "Transcript of r2P6c5tzvM8 (2325 chars):
   If you're Canadian, keep watching. It's election time. I've just requested that the Governor General dissolve Parliament and call it an election, and she has agreed. The campaign will last 36 days, and the election will be on April 28th. There are 343 ridings up for grabs across the country. ... Here's what the House looks like right now with the Liberal minority government. The Liberals have 152 seats. The Conservatives have 120. The Bloc Quebecois have 33. The NDP have 24. The Green Party has 2. Three MPs are sitting as independents, and four seats are vacant. For a majority government, a party needs to win at least 172 seats, but a minority government can be formed with fewer than that. ..."

[OBSERVE]:
  TOOL CALLS (1):
    OK  transcribe({'video_path': 'r2P6c5tzvM8'})  ->  32.2s

[JUDGE] iter 16 (model: qwen/qwen3.5-9b):
  Reasoning 4 | Action 5 | Granularity 5 | Progress 4
  Redundancy 5 | Think 5 | Code 5 | Output 5 | Tool 5
  WEIGHTED TOTAL: 60.5/62.5 (96.8%)
\end{agentlog}

\noindent\textbf{Observation.} 
When root gets conflicting visual information, it decides to call \texttt{transcribe} to use transcript evidence for cross-modal conflict resolution.

\subsection{Statistics}

\paragraph{Tool Usage.}
As shown in \autoref{tab:rlm-tool-calls}, average tool usage indicates a strongly adaptive exploration strategy. In practice, the agent schedules all tools in 14 of 19 queries. In practice, it uses Caption tool ($10\pm3$) mainly as a relevance filter, and relies more on targeted extraction with QueryClaim tool($8\pm2$) than on broader summarization with General Notes tool ($4\pm2$). Transcribe tool is used only as a supplement, while Retrieval tool is rarely needed ($1\pm1$), since most videos are short ($<60$s) and do not require fine-grained temporal localization.

\begin{table}[t]
\centering
\small
\setlength{\tabcolsep}{3pt}
\begin{tabular}{lcccccc c}
\toprule
\multicolumn{8}{c}{\textbf{Tool Calls (Average)}} \\
 & Capt. & Claims & Notes & QA & Trans. & Retr. & Total \\
\midrule
Q1 & 6 & 6 & 4 & 6 & 4 & 1 & 28 \\
Q2 & 6 & 6 & 4 & 6 & 2 & 2 & 26 \\
Q3 & 5 & 6 & 5 & 4 & 2 & 2 & 24 \\
Q4 & 5 & 5 & 2 & 2 & 2 & 0 & 15 \\
Q5 & 8 & 6 & 2 & 2 & 1 & 1 & 21 \\
Q6 & 7 & 10 & 4 & 4 & 2 & 2 & 29 \\
Q7 & 11 & 12 & 3 & 4 & 1 & 1 & 32 \\
Q8 & 10 & 10 & 4 & 2 & 2 & 1 & 28 \\
Q9 & 7 & 7 & 1 & 1 & 0 & 0 & 16 \\
Q10 & 17 & 5 & 3 & 4 & 0 & 1 & 30 \\
Q11 & 18 & 10 & 2 & 2 & 0 & 0 & 32 \\
Q12 & 11 & 8 & 7 & 2 & 3 & 2 & 33 \\
Q13 & 10 & 5 & 6 & 2 & 2 & 0 & 24 \\
Q14 & 10 & 7 & 3 & 2 & 1 & 1 & 24 \\
Q15 & 10 & 8 & 4 & 3 & 2 & 1 & 28 \\
Q16 & 10 & 10 & 4 & 1 & 1 & 1 & 26 \\
Q17 & 11 & 12 & 6 & 2 & 1 & 0 & 32 \\
Q18 & 10 & 10 & 2 & 1 & 1 & 2 & 26 \\
Q19 & 10 & 10 & 1 & 1 & 1 & 1 & 24 \\
\midrule
Avg. & 10$\pm$3 & 8$\pm$2 & 4$\pm$2 & 3$\pm$2 & 2$\pm$1 & 1$\pm$1 & 26$\pm$5 \\
\bottomrule
\end{tabular}
\caption{Statistics of tool calls per query (averaged over two runs). }
\label{tab:rlm-tool-calls}
\end{table}

\paragraph{Runtime Performance.}
The evaluation by LLM-as-a-judge indicates strong behavioral quality (92$\pm$2\% on average), with near-perfect Output Waste and Code Minimality scores as reported in \autoref{tab:rlm-query}.
The average wall time per query is 36$\pm$12 minutes, with most of the runtime spent on tool execution rather than \rootlm{} inference.

\begin{table}[t]
\centering
\small
\setlength{\tabcolsep}{3pt}
\begin{tabular}{l cc ccc}
\toprule
 & \multicolumn{2}{c}{\textbf{Quality}} & \multicolumn{3}{c}{\textbf{Latency}} \\
\cmidrule(lr){2-3} \cmidrule(lr){4-6}
 & Judge (\%) & Facts/Iter & Wall (min) & Tokens (K) & Root LLM (s) \\
\midrule
Q1 & 88 & 0.74 & 46 & 595 & 382 \\
Q2 & 92 & 0.57 & 47 & 510 & 326 \\
Q3 & 86 & 0.90 & 30 & 516 & 311 \\
Q4 & 94 & 1.85 & 24 & 329 & 216 \\
Q5 & 95 & 1.11 & 36 & 405 & 278 \\
Q6 & 92 & 0.37 & 49 & 476 & 342 \\
Q7 & 92 & 1.08 & 42 & 493 & 416 \\
Q8 & 89 & 1.72 & 48 & 563 & 409 \\
Q9 & 91 & 1.39 & 16 & 175 & 199 \\
Q10 & 93 & 0.99 & 26 & 418 & 357 \\
Q11 & 93 & 1.56 & 22 & 430 & 300 \\
Q12 & 93 & 1.28 & 22 & 472 & 481 \\
Q13 & 92 & 1.00 & 22 & 271 & 346 \\
Q14 & 92 & 0.93 & 36 & 457 & 366 \\
Q15 & 89 & 1.21 & 45 & 470 & 351 \\
Q16 & 92 & 0.86 & 48 & 613 & 342 \\
Q17 & 90 & 0.81 & 54 & 419 & 298 \\
Q18 & 94 & 0.91 & 37 & 378 & 235 \\
Q19 & 95 & 1.26 & 26 & 329 & 244 \\
\midrule
Avg. & 92$\pm$2 & 1.1$\pm$0.4 & 36$\pm$12 & 438$\pm$109 & 326$\pm$72 \\
\bottomrule
\end{tabular}%
\caption{Statistics of behavior quality and query latency, averaged over two runs.}
\label{tab:rlm-query}
\end{table}

\paragraph{\rootlm{} Context Window Usage}
As detailed in Table \ref{tab:rlm-token}, \rootlm{} use an average of only 33\% of its 32K context window, avoiding frequent truncation. The context growth scales linearly to roughly 1.3\% per iteration (from 8\% at Iteration 0 to 61\% at Iteration 40). This stable progression indicates that the accumulation of both the Memory Bank and history is  controllable, therefore preventing explosive token consumption in later iterations.

\begin{table}[t]
\centering
\label{tab:token-usage}
\setlength{\tabcolsep}{4pt}
\small
\begin{tabular}{lrl}
\toprule
\textbf{Metric} & \textbf{Value} & \\
\midrule
\multicolumn{3}{l}{\textit{Per-Iteration Token Consumption}} \\
\quad Total tokens & 11131 & \\
\quad Prompt (context) & 10689 $\pm$ 6434 & (96\%) \\
\quad Completion & 442 $\pm$ 415 & (4\%) \\
\quad \quad Reasoning & 328 & \\
\quad \quad Output & 114 & \\
\midrule
\multicolumn{3}{l}{\textit{Context Window Utilization (32K limit)}} \\
\quad Average usage & 33\% $\pm$ 20\% & \\
\quad Maximum usage & 97\% & \\
\quad Iterations $>$80\% & 25 & (1.7\%) \\
\midrule
\multicolumn{3}{l}{\textit{Context Growth Over Iterations}} \\
\quad Iteration 0 & 2,582 & (8\%) \\
\quad Iteration 20 & 9,983 & (31\%) \\
\quad Iteration 40 & 20,026 & (61\%) \\
\bottomrule
\end{tabular}
\caption{Root LM context-window usage, averaged over two runs.}
\label{tab:rlm-token}
\end{table}

\section{Appendix: Prompt Templates}

\label{app:prompts}





This appendix collects all prompt templates used by the system. Prompts are grouped by the method component that uses them. Implementation details for each method are provided in Appendices~\ref{app:retrieval}--\ref{app:rlm}.

\subsection{Retrieval Query Expansion Prompt}
\label{app:retrieval-query-expansion-prompt}

\paragraph{Sub-query decomposition prompt.}
The decomposition prompt consumes the structured fields of a MAGMaR query and
emits a JSON array of fine-grained search phrases. It is used with
Qwen3.5-27B and thinking disabled.

\begin{promptbox}
\pline{You are a research decomposition specialist. Your task is to take a user's query}
\pline{and break it down into an exhaustive set of searchable sub-queries -- short}
\pline{phrases or keyword combinations that could be entered into a search engine or}
\pline{database to retrieve all the information needed to fully answer the original query.}
\pline{}
\pline{You will receive the following inputs:}
\pline{- Title: \textless TITLE\textgreater}
\pline{- Language: \textless LANGUAGE\textgreater}
\pline{- Persona: \textless PERSONA\_TITLE\textgreater}
\pline{- Background: \textless BACKGROUND\textgreater}
\pline{- Query: \textless QUERY\textgreater}
\pline{}
\pline{Decomposition Rules:}
\pline{1. Coverage: Extract every distinct piece of information the user is asking for. Do not merge separate information needs into one sub-query.}
\pline{2. Granularity: Each sub-query should target ONE specific, retrievable piece of information. Prefer atomic queries over compound ones.}
\pline{3. Implicit needs: Go beyond what is explicitly stated. Based on the background and persona\_title, infer what additional information the user would likely need but did not explicitly ask for.}
\pline{4. Search-friendly format: Each sub-query should be phrased as a concise search phrase, typically 3--10 words, not a full sentence or question.}
\pline{5. Context anchoring: Each sub-query should include enough context to be independently searchable without ambiguity.}
\pline{6. Source-awareness: If the user requests source information, generate sub-queries targeting official sources, methodologies, and data provenance.}
\pline{7. Dimensional expansion: Consider additional perspectives or breakdowns by time, place, category, cause, mechanism, or comparison only when they add value.}
\pline{8. No redundancy: Each sub-query must be meaningfully distinct.}
\pline{9. Language: Always generate sub-queries in English.}
\pline{10. Generate between 10 and 25 sub-queries.}
\pline{11. Do not mechanically prepend the full topic title to every sub-query.}
\pline{12. Focus on the specific information being sought, not on repeating the topic name.}
\pline{}
\pline{Return ONLY a JSON array of strings. No explanation, no markdown, no code blocks.}
\end{promptbox}











































\paragraph{General note extraction prompt.}
The general-note prompt is query-agnostic but not fully context-free: it includes the source topic and video identifier together with an evidence-first instruction block.

\begin{promptbox}
\pline{You are extracting observation notes directly from a raw video.}
\pline{}
\pline{Video context:}
\pline{- topic: \textless TOPIC\textgreater}
\pline{- video\_id: \textless VIDEO\_ID\textgreater}
\pline{- timestamp\_span: \textless TIMESTAMP\_SPAN\_OR\_NULL\textgreater}
\pline{}
\pline{Rules:}
\pline{- Record only directly observable content.}
\pline{- No inference, speculation, causality, or cross-video synthesis.}
\pline{- Capture OCR (on-screen text), events, and visible scene details.}
\pline{- One note per atomic visible, audible, or textual fact.}
\pline{- Use modality `visual' for scene content, `ocr' for on-screen text, and `audio' for transcript or speech.}
\pline{- Use the provided timestamp span for each note when no narrower timestamp is available.}
\pline{- If there is no usable evidence, return an empty notes list.}
\pline{}
\pline{Output strict JSON only.}
\pline{No markdown, no code fences, no explanation, no extra keys outside the schema.}
\pline{Expected shape:}
\pline{\{}
\pline{\ \ "notes": [}
\pline{\ \ \ \ \{}
\pline{\ \ \ \ \ \ "text": "...",}
\pline{\ \ \ \ \ \ "modality": "visual",}
\pline{\ \ \ \ \ \ "timestamp": [0.0, 1.0]}
\pline{\ \ \ \ \}}
\pline{\ \ ]}
\pline{\}}
\end{promptbox}

\paragraph{Query-conditioned claim extraction prompt.}
The single-query claim-extraction prompt conditions on the query text together with persona, background, topic, and video identity.

\begin{promptbox}
\pline{You are extracting query-relevant claims directly from a raw video.}
\pline{}
\pline{Query context:}
\pline{- query\_id: \textless QUERY\_ID\textgreater}
\pline{- topic: \textless TOPIC\textgreater}
\pline{- persona\_title: \textless PERSONA\_TITLE\textgreater}
\pline{- background: \textless BACKGROUND\textgreater}
\pline{- query: \textless QUERY\_TEXT\textgreater}
\pline{- video\_id: \textless VIDEO\_ID\textgreater}
\pline{}
\pline{Rules:}
\pline{- Extract up to \textless PER\_VIDEO\_TARGET\textgreater\ claims relevant to the query from this video.}
\pline{- Claims must be directly supported by observable video content.}
\pline{- Avoid generic scene summary unless it directly serves the query.}
\pline{- Avoid duplicates and paraphrases.}
\pline{- If the video does not contain evidence for the query, return an empty claims list.}
\pline{- source must be one of `video\_visual', `video\_text', or `transcript'.}
\pline{- timestamp must be [start, end].}
\pline{- confidence must be a float between 0 and 1.}
\pline{}
\pline{Output strict JSON only.}
\pline{No markdown, no code fences, no explanation, no extra keys outside the schema.}
\pline{Expected shape:}
\pline{\{}
\pline{\ \ "claims": [}
\pline{\ \ \ \ \{}
\pline{\ \ \ \ \ \ "claim": "...",}
\pline{\ \ \ \ \ \ "confidence": 0.85,}
\pline{\ \ \ \ \ \ "evidence": "...",}
\pline{\ \ \ \ \ \ "source": "video\_visual",}
\pline{\ \ \ \ \ \ "timestamp": [0.0, 1.0]}
\pline{\ \ \ \ \}}
\pline{\ \ ]}
\pline{\}}
\end{promptbox}

\paragraph{Decompose Query into Questions. } Expand single query into a series of question--answer pairs based on a provided title, language, persona, and background.
\begin{promptbox}
You are a research decomposition specialist. Your task is to take a user's query and break it down into an exhaustive set of searchable research questions — complete questions that could be used to retrieve all the information needed to fully answer the original query.

You will receive the following inputs:

- Title: \{title\}

- Language: \{language\}

- Persona: \{persona\_title\}

- Background: \{background\}

- Query: \{query\}

Decomposition Rules:

1. Coverage: Extract every distinct piece of information the user is asking for. Do not merge separate information needs into one question. If the query asks for multiple related but distinct data points, each one should become its own question.

2. Granularity: Each question should target ONE specific, retrievable piece of information. Prefer atomic questions over compound ones.

3. Implicit needs: Go beyond what is explicitly stated. Based on the background and persona\_title, infer what additional information the user would likely need but did not explicitly ask for. Consider what a professional in that role would typically require to produce complete, high-quality work on this topic.

4. Search-friendly format: Each sub-query must be written as a concise, well-formed question that could plausibly be entered into a search engine or research database.

5. Context anchoring: Each question should include enough context (e.g., specific names, dates, locations, technical terms) to be independently searchable without ambiguity.

6. Source-awareness: If the user requests source information or credibility indicators, generate questions specifically targeting official sources, methodologies, and data provenance.

7. Dimensional expansion: For each core information need identified, consider whether the user would benefit from additional perspectives or breakdowns. Ask yourself: can this information be meaningfully decomposed further by time, place, category, cause, mechanism, comparison, or any other axis that is natural and relevant to the topic? Only expand along dimensions that genuinely add value given the query's subject matter and the user's background.

8. No redundancy: Each question must be meaningfully distinct. Do not produce near-duplicates that would return the same search results.

9. Language: Always generate questions in English, regardless of the language field in the input.

10. Quantity: Generate between 10 and 25 questions. Focus on quality and relevance over quantity.

11. Avoid mechanical repetition: Do not mechanically prepend the full topic title to every question. Each question should contain only the context necessary for an effective search.

12. Focus on information needs: Focus on the specific information being sought rather than repeating the topic name unnecessarily.

Return ONLY a JSON array of strings. No explanation, no markdown, and no code blocks.

For example, given a query about the 2025 Canadian federal election asking for seat counts and vote shares, good questions would be:

[
"What was the total number of seats won by each political party in the 2025 Canadian federal election?",
"What percentage of the national popular vote did each major party receive in the 2025 Canadian federal election?",
"How many seats did each party gain or lose compared with the 2021 Canadian federal election?",
"What official datasets published by Elections Canada contain vote totals and seat counts for the 2025 federal election?",
"What demographic voting patterns were observed in the 2025 Canadian federal election?"
]

NOT:

[
"What happened in the 2025 Canadian federal election?",
"What were the results of the 2025 Canadian federal election?",
"What information is available about the 2025 Canadian federal election?"
]

JSON array:
\end{promptbox}

\paragraph{Question Answering. } Qwen 3.5 produces an answer to the question based on the down-sampled video and transcript.
\begin{promptbox}
Question:
{question}

Answer concisely using the video and transcript. Answer to the best of your abilities. If you can't answer all of the question, answer the parts that you can (Ex. if asked about Liberal and Conservative vote counts, but only have the Liberal vote counts, return those). If you have no information about anything related to the question, return "I don't know". Never say the event hasn't taken place.

Return ONLY the final factual answer.
\end{promptbox}

\paragraph{Combined Answers. } Combines all the answers generated independently from each relevant video in the top k most similar.
\begin{promptbox}
You are given extracted answers from videos. These answers are factual and must be used.
        
Question: \{subquery\}

Extracted Answers (treat as ground truth):
\{valid\_answers\}

Combine them into a single answer.

Do NOT use prior knowledge.
Do NOT say the event has not taken place.
Only use the provided answers.

If you recieve conflicting information, make a best guess NOT on prior knowlege but based on the nature of the question (Ex. for a question about how many seats a party has one, its reasonable to assume the largest number is the most recent seat count)

Return only the final answer. You might not be able to answer it fully, and that's okay. Answer what you can and then say specifically what information is unknown.

\end{promptbox}

\paragraph{Generate Follow Up Question. } Generates the follow up question in the iterative QA system based on entire past context.
\begin{promptbox}
You are refining a research question based on prior answers.

Context:
{context}

Generate ONE new question that:
- extracts new information not yet covered
- is more specific or differently framed

If no meaningful new question can be formed, output:
NONE

Return ONLY the question or NONE.
\end{promptbox}

\paragraph{Qwen 3.5 calibration prompt.}
The prompted Qwen 3.5 backend receives the full source video together with one extracted artifact and is instructed to return a scalar probability in a constrained answer format.

\begin{promptbox}
\pline{To help you make more accurate and consistent judgments, here is an expanded explanation of how to interpret and assign support percentages.}
\pline{These examples are designed to cover a range of real-world cases you may encounter in the annotation task.}
\pline{100\% - /Full and unambiguous support:}
\pline{The video clearly shows the exact event described in the claim. There is no need for guessing or interpretation.}
\pline{80-100\% - Almost complete support:}
\pline{The main content in the claim is shown, but there may be minor ambiguity in location, identity, or completeness. The overall claims are supported by the video.}
\pline{60-80\% - Strong partial support:}
\pline{The video strongly suggests the claim is true, but some critical details may be missing, obscured, or ambiguous, limiting the ability to confirm the claim with certainty. The video gives strong but not definitive support.}
\pline{40-60\% - Moderate partial support:}
\pline{There is some alignment with the claim, but large portions are either missing, unclear, or open to interpretation. While the footage may point in the same general direction as the claim, it lacks the clarity or completeness needed for confident verification.}
\pline{20-40\% - Minimal weak support:}
\pline{There are small visual or audio cues that could hint at the claim, but they are insufficient to be confident.}
\pline{0-20\% - Very weak or speculative support:}
\pline{There may be the slightest indirect reference, such as a related object or setting, but nothing concrete happens.}
\pline{0\% - No support or contradiction:}
\pline{The video does not relate to the claim at all, or it directly shows something opposite.}
\pline{}
\pline{Based on the provided video and text, evaluate the probability that the text is true.}
\pline{Your answer must be a decimal number between 0 and 1, and you must strictly follow the format below:}
\pline{\textless answer\textgreater probability\_value\textless /answer\textgreater}
\pline{Where probability\_value is the result you calculate.}
\pline{The text to evaluate is:}
\pline{}
\pline{\textless ARTIFACT\_TEXT\textgreater}
\end{promptbox}

Malformed outputs trigger a stricter retry prompt that preserves the same task content while requiring a single answer in the exact form \texttt{<answer>0.73</answer>}.

\paragraph{Baseline}
The baseline approach feeds all query-conditioned claims into a single LLM
prompt and generates the complete report in one forward pass.

\begin{promptbox}
\pline{You are a report writing assistant. Your task is to synthesize a set of claims extracted from multiple videos into a single, fluent, well-organized report that answers the given query.}
\pline{}
\pline{\#\# Instructions:}
\pline{1. Read all the claims below carefully. Each claim was extracted from a specific video and has a timestamp.}
\pline{2. Group related claims together logically (e.g., by sub-topic or chronological order).}
\pline{3. Write a coherent, well-structured report that covers all the key information from the claims.}
\pline{4. For EVERY piece of information in your report, include an inline citation in the format [video\_id, timestamp\_start-timestamp\_end].}
\pline{5. If multiple claims from different videos support the same point, cite all relevant sources.}
\pline{6. Remove redundant information --- if multiple claims say the same thing, mention it once and cite all sources.}
\pline{7. The report should be fluent and readable, not a list of bullet points.}
\pline{8. Keep the report concise but comprehensive (aim for 200-400 words).}
\pline{}
\pline{\#\# Query/Topic: \{topic\}}
\pline{}
\pline{\#\# Claims:}
\pline{\{claims\_text\}}
\pline{}
\pline{\#\# Report:}
\end{promptbox}

\paragraph{\ginger{} clustering prompt.}
The model receives all claims for a query and is instructed to partition
them into thematic facet clusters, returning a labeled JSON partition of
the claim set.

\begin{promptbox}
\pline{You are an information analyst. Given a set of claims about a topic extracted from videos, group them into distinct facet clusters. Each cluster should represent a different sub-topic or aspect of the main topic.}
\pline{}
\pline{\#\# Instructions:}
\pline{1. Read all claims carefully.}
\pline{2. Group them into clusters based on their sub-topic/facet (e.g., "casualties", "rescue efforts", "damage assessment", "government response", etc.).}
\pline{3. Each claim should belong to exactly one cluster.}
\pline{4. Give each cluster a short, descriptive label.}
\pline{5. Output your result as a JSON object with the following format:}
\pline{}
\pline{\{}
\pline{  "clusters": [}
\pline{    \{}
\pline{      "label": "Short descriptive label for this facet",}
\pline{      "claim\_ids": ["qc-10-xxx-000", "qc-10-xxx-001"]}
\pline{    \},}
\pline{    ...}
\pline{  ]}
\pline{\}}
\pline{}
\pline{Only output the JSON object, no other text.}
\pline{}
\pline{\#\# Topic: \{topic\}}
\pline{}
\pline{\#\# Claims:}
\pline{\{claims\_text\}}
\end{promptbox}

\paragraph{\ginger{} ranking prompt.}
The model receives the labeled clusters and is instructed to rank them by
relevance to the query topic, returning an ordered JSON array of cluster
labels.

\begin{promptbox}
\pline{You are a relevance assessor. Given a query/topic and a list of facet clusters (each containing grouped claims from videos), rank the clusters from most to least relevant to the query.}
\pline{}
\pline{\#\# Instructions:}
\pline{1. Consider which facets are most important for answering/addressing the query topic.}
\pline{2. Rank all clusters from most relevant to least relevant.}
\pline{3. Output a JSON array of cluster labels in order from most to least relevant:}
\pline{}
\pline{\{}
\pline{  "ranked\_labels": ["most relevant label", "second most relevant", ...]}
\pline{\}}
\pline{}
\pline{Only output the JSON object, no other text.}
\pline{}
\pline{\#\# Topic: \{topic\}}
\pline{}
\pline{\#\# Clusters:}
\pline{\{clusters\_text\}}
\end{promptbox}

\paragraph{\ginger{} summarization prompt.}
The model receives the claims within a single cluster and is instructed
to condense them into one cited sentence of at most 40 words, preserving
inline citations anchored to the supporting evidence.

\begin{promptbox}
\pline{You are a concise summarizer. Summarize the following cluster of claims into a SINGLE sentence (maximum 40 words). The sentence must:}
\pline{1. Capture the key information from all claims in this cluster.}
\pline{2. Include inline citations in the format [video\_id, timestamp] for every fact mentioned.}
\pline{3. Be factual --- only include information present in the claims.}
\pline{}
\pline{\#\# Cluster: \{cluster\_label\}}
\pline{}
\pline{\#\# Claims in this cluster:}
\pline{\{cluster\_claims\_text\}}
\pline{}
\pline{\#\# One-sentence summary:}
\end{promptbox}

\paragraph{\ginger{} fluency prompt.}
The model receives the concatenated one-sentence cluster summaries and is
instructed to rewrite them into a coherent 200--400-word prose report
without adding new information or removing any citations.

\begin{promptbox}
\pline{You are an editor. Below is a report composed of individual summary sentences about the topic "\{topic\}". Your task is to rewrite it into a smooth, fluent, well-organized report.}
\pline{}
\pline{\#\# Rules:}
\pline{1. Do NOT add any new information that is not in the summaries below.}
\pline{2. Do NOT remove any information or citations from the summaries.}
\pline{3. Keep ALL inline citations in the format [video\_id, timestamp].}
\pline{4. Improve transitions between sentences for better readability.}
\pline{5. You may reorder sentences for better logical flow.}
\pline{6. Keep the report concise (200-400 words).}
\pline{}
\pline{\#\# Draft report (concatenated summaries):}
\pline{\{draft\_report\}}
\pline{}
\pline{\#\# Final polished report:}
\end{promptbox}

\paragraph{\marquis-\rlm{} REPL system prompt.}
\begin{promptbox}
\pline{You answer queries using an interactive Python REPL, called iteratively until you submit a final answer.}
\pline{}
\pline{THINK-ACT-OBSERVE LOOP:}
\pline{  Each iteration: THINK (brief reasoning), ACT (one code block), OBSERVE the output.}
\pline{  THINK phase: READ the memory snapshot below --- it shows your findings (global knowledge) and per-video facts. Base your next action on what you ALREADY KNOW, not assumptions.}
\pline{}
\pline{\{pacing\}}
\pline{}
\pline{ENVIRONMENT:}
\pline{  - context['task'], context['video\_pool'], context['tools'] are read-only.}
\pline{  - `memory` is a persistent dict (survives compaction).}
\pline{  - Tools are pre-loaded as plain Python functions; call them directly.}
\pline{}
\pline{}
\pline{FORMAT: THINK (2-4 sentences), then ONE ```repl``` code block (1-5 lines, ONE tool call). NO for-loops over videos.}
\pline{}
\pline{FINAL ANSWER: report = write\_report(memory['selected\_facts']), then FINAL\_VAR(report) outside the code block.}
\end{promptbox}

\paragraph{\marquis-\rlm{} \rootlm{} Think prompt.}

\begin{promptbox}
\pline{TASK: \{query\_text\}}
\pline{}
\pline{CURRENT FINDINGS:}
\pline{\{findings\_str\}}
\pline{}
\pline{FACT TABLE SUMMARY:}
\pline{\{fact\_summary\}}
\pline{}
\pline{VIDEO STATUS:}
\pline{\{video\_status\}}
\pline{}
\pline{You are the analytical brain. Based on all facts collected so far:}
\pline{}
\pline{1. NEW\_FINDINGS: List any new high-level findings (one sentence each) not already in CURRENT FINDINGS. If a new fact CONTRADICTS an existing finding, say CONFLICT: <existing> vs <new>.}
\pline{}
\pline{2. UPDATED\_FINDINGS: Output the complete updated findings list (old + new, deduplicated). One finding per line, prefixed with `- '.}
\pline{}
\pline{3. NEXT\_STEPS: What should the agent do next? Be specific: which video, which tool, which question.}
\pline{}
\pline{Be concise.}
\end{promptbox}

\paragraph{\marquis-\rlm{} \rootlm{} Judge prompt.}

\begin{promptbox}
\pline{TASK: \{query\_text\}}
\pline{}
\pline{FINDINGS (root's current understanding):}
\pline{\{findings\_str\}}
\pline{}
\pline{FACT TABLE (\{n\} facts):}
\pline{\{fact\_lines\}}
\pline{}
\pline{You are a strict quality judge. Review ALL facts above for the task.}
\pline{}
\pline{1. ITEM REVIEW: For each fact (F\#0, F\#1, ...), give a verdict.}
\pline{   BE CONSERVATIVE --- only REMOVE if clearly irrelevant or duplicate. When in doubt, KEEP.}
\pline{     KEEP    --- useful, specific, or even mildly relevant (default)}
\pline{     REMOVE  --- clearly irrelevant or duplicate of another listed fact}
\pline{     REWRITE --- needs more detail or has a missing timestamp (flag, do NOT drop)}
\pline{   Format: F\#0: KEEP / F\#3: REMOVE (dup of F\#1) / F\#5: REWRITE (missing timestamp)}
\pline{}
\pline{2. SELECTED: Pick the 10-40 BEST facts for a comprehensive report (prefer MORE coverage). List their IDs: SELECTED: F\#0, F\#2, F\#7, ...}
\pline{}
\pline{3. MISSING TIMESTAMPS: List facts that are useful but lack timestamps; suggest video\_qa queries to resolve them.}
\pline{}
\pline{4. GAPS: What information is still missing for a thorough report?}
\pline{}
\pline{5. READY: Can we write a good report now? (yes / no / almost)}
\pline{}
\pline{Be specific and concise.}
\end{promptbox}

\paragraph{\marquis-\rlm{} LLM-as-a judge prompt (behavior-level).}

\begin{promptbox}
\pline{You are evaluating an AI agent's performance on iteration \{iteration\}/\{max\_iter\}.}
\pline{TASK:    \{query\}}
\pline{MEMORY STATE BEFORE: \{mem\_before\}}
\pline{THINK:   \{think\_text\}}
\pline{ACT:     \{code\}}
\pline{OBSERVE: \{observe\}}
\pline{MEMORY STATE AFTER:  \{mem\_after\}}
\pline{}
\pline{Rate each dimension 1-5 with ONE sentence justification.}
\pline{}
\pline{\#\# Core dimensions:}
\pline{1. Reasoning (1-5):   Did THINK show sound reasoning based on memory?}
\pline{2. Action (1-5):      Was the chosen action relevant and logical?}
\pline{3. Granularity (1-5): One focused step, or too much at once?}
\pline{4. Progress (1-5):    Did this iteration meaningfully advance the task?}
\pline{}
\pline{\#\# Efficiency breakdown (5 sub-scores):}
\pline{5a. Eff\_Redundancy (1-5)         --- avoided repeating a tool call?}
\pline{5b. Eff\_Think\_Conciseness (1-5) --- THINK tight and non-repetitive?}
\pline{5c. Eff\_Code\_Minimality (1-5)   --- minimal code for its purpose?}
\pline{5d. Eff\_Output\_Waste (1-5)      --- avoided producing useless output?}
\pline{5e. Eff\_Tool\_Choice (1-5)       --- most cost-effective tool for this sub-goal?}
\pline{}
\pline{Format EXACTLY: one line per dimension as `Name: <score> --- <reason>', then `TOTAL: <sum>/45'.}
\end{promptbox}

\end{document}